\documentclass{aastex} 

\shorttitle{The nitrogen abundance of massive stars}

\shortauthors{C.~Aerts et al.}

\begin{document} 

\title{The surface nitrogen abundance of a massive star in relation to its
  oscillations, rotation, and magnetic field}

\author{C.~Aerts\altaffilmark{1,2,3}, G.~Molenberghs\altaffilmark{3,4},
M.~G.\ Kenward\altaffilmark{5} \and C.~Neiner\altaffilmark{6}}

\altaffiltext{1}{Institute of Astronomy, KU\,Leuven, Celestijnenlaan 200D,
B-3001 Leuven, Belgium}
\altaffiltext{2}{Department of Astrophysics, IMAPP, Radboud University
Nijmegen, P.O.\ Box 9010, 6500 GL Nijmegen, The Netherlands}
\altaffiltext{3}{Faculty of Science, Hasselt University, Martelarenlaan 42,
B-3500 Hasselt,  Belgium}
\altaffiltext{4}{I-BioStat, KU\,Leuven, Kapucijnenvoer 35, 
B-3000 Leuven, Belgium}
\altaffiltext{5}{Department of Medical Statistics, London School of Hygiene and
  Tropical Medicine, Keppel Street, 
London WC1E7HT, United Kingdom}
\altaffiltext{6}{LESIA, UMR 8109 du CNRS, Observatoire de Paris, UPMC, Paris
  Diderot, 5 Place Jules Janssen 92195 Meudon Cedex, France}

\begin{abstract}
We have composed a sample of 68 massive stars in our galaxy whose projected
rotational velocity, effective temperature and gravity are available from
high-precision spectroscopic measurements. The additional seven observed
variables considered here are their surface nitrogen abundance, rotational
frequency, magnetic field strength, and the amplitude and frequency of their
dominant acoustic and gravity mode of oscillation.  Multiple linear regression
to estimate the nitrogen abundance combined with principal components analysis,
after addressing the incomplete and truncated nature of the data, reveals that
the effective temperature and the frequency of the dominant acoustic oscillation
mode are the only two significant predictors for the nitrogen abundance, 
  while the projected rotational velocity and the rotational frequency have no
  predictive power.  The dominant gravity mode and the magnetic field strength
are correlated with the effective temperature but have no predictive power for
the nitrogen abundance.  
Our findings are completely based on observations and their proper
  statistical treatment and call for a new strategy in evaluating the
  outcome of stellar evolution computations.

\end{abstract}

\keywords{methods: statistical -- stars:abundances --
stars:  massive -- stars: magnetic field -- 
stars: oscillations (including pulsations)}

\section{Introduction}

Mixing of chemical species inside stars is poorly understood. Yet, its effect on
stellar evolution and supernova explosions, and by implication on the chemical
enrichment of galaxies is of prime importance.  It was realised already
  quite some time ago that stellar rotation and the mixing it induces must play
  a major role in stellar evolution theory \citep[e.g.,][]{Herrero1992} but its
  inclusion in models is far from trivial, see e.g.,
  \citet{Langer1992,Talon1997} for early attempts and discussions, while
  \citet{Maeder2009} is a recent extensive monograph in this topic.

Ways to test the theoretical descriptions used to represent rotational mixing
are scarce and are mainly limited to studies of the chemical composition and
projected rotational velocity of stellar atmospheres \citep[e.g.,][among many
  others]{Venn2005,Hunter2008,Przybilla2010}.  A particularly strong
observational diagnostic for mixing is the surface nitrogen abundance of stars
that are undergoing hydrogen fusion through the carbon-nitrogen-oxygen cycle in
their interior. It turns out that the current theoretical concepts of rotational
mixing seemingly fail to explain observations of various massive stars in the
Milky Way and in the Magellanic Clouds  \citep{Hunter2008,Brott2011,
    Rivero2012,Bouret2012,Bouret2013},  leading to intense debates on yet
unknown causes of the discrepancies and on the way to identify other
physical ingredients lacking in the theoretical models
\citep[e.g.,][]{Meynet2011,Potter2012,Mathis2013}.

Here, we shed new light on the matter by considering a variety of {\it
  observational\/} data and by subjecting them to careful statistical
analysis. In particular, we investigate the relationship between observed
stellar oscillations, rotational frequency, magnetic field strength and surface
nitrogen abundance in a sample of galactic massive stars for which this
multitude of data has recently become available.   A direct comparison
  between observed rotational properties with evolutionary models would ideally
  be based on a value for the equatorial rotation velocity. This would require
  having either precise measurements of both the rotational frequency and the
  stellar radius, or else of the inclination angle $i$ between the rotational
  axis of the star and the line-of-sight of the observer, in addition to
  $v\sin\,i$. While the combination $(v\sin\,i,i)$ is available for some
  pulsating stars from an analysis of time-series spectroscopy and for a few
  magnetic stars from spectro-polarimetry, this concerns very few
  stars. Similarly, direct measurements of the radii are in general not
  available for massive stars. Hence, a specific point of attention in this
  work is whether
  the use of the rotational frequency, rather than the projected rotational
  velocity, leads to an improved diagnostic.

\section{Sample Selection}

\begin{figure}[t!]
\begin{center}
\rotatebox{270}{\resizebox{8cm}{!}{\includegraphics{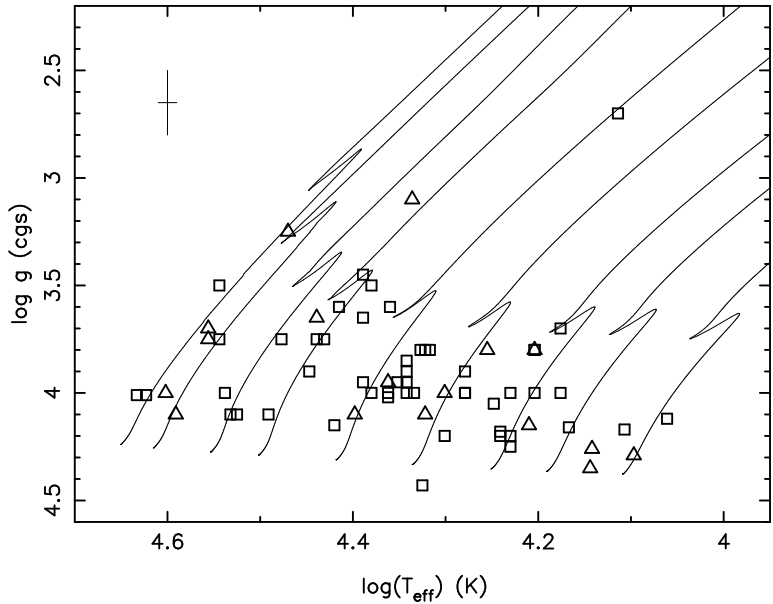}}}
\end{center}
\caption[]{ A ($\log\,T_{\rm eff}, \log\,g$) diagram of the measured stars
  (squares for single stars, triangles for spectroscopic binaries) and of
  evolutionary tracks (full lines, see text) for masses of 3, 4, 5, 7, 10, 15,
  20, 30, 40\,M$_\odot$, from right to left. A typical error box is indicated in
  the upper left corner.}
\label{fig1}
\end{figure}

Sample selection was restricted to stars whose effective temperature, gravity
and projected rotational velocity are available from high signal-to-noise
high-resolution spectroscopy.  From this sample, we kept the stars for which at
most one of the following four additional properties has missing data: nitrogen
abundance, rotational frequency, magnetic field strength, and oscillations.
 This rather strict criterion of missingness was adopted to achieve a good
  starting point for the statistical analysis outlined in the following section
  and implied that only Galactic OB stars were retained in the sample.

Regarding the oscillations detected,  we considered the frequency and the
amplitude of the dominant acoustic mode and of the dominant gravity mode of the
star \citep[see][for a definition and extensive description of such
    heat-driven oscillations in massive stars]{Aerts2010} whenever available in
  the literature.  We downloaded the Hipparcos data from \citet{Perryman1997}
  and derived upper limits for the oscillation amplitudes as four times the
standard deviation of these data. These upper limits were adopted in the
case of stars for which one or both types of oscillations have not yet been
detected.

The magnetic field strength under the assumption of an oblique dipole
  configuration is either available as a measured value (in case a time series
  of spectro-polarimetry was observed) or as a lower limit (because the
  projection factor between the line-of-sight and the magnetic axis at the time
  of measurement is unknown).  We adopted the most recent published results
  derived from high-resolution polarimeters such as ESPaDOnS, NARVAL or HARPSpol
  (data sources are listed along with the data in Table\,\ref{data}).  In
  absence of those, we took results based on the lower-resolution spectrographs
  such as, e.g., FORS1/2. For a comparison between the interpretations based on
  these two types of measurements, we refer to \citet{Shultz2012}. We added to
  those data all unpublished null detections obtained by the MiMeS (Magnetism in
  Massive Stars) 
  collaboration\footnote{{\tt
    http://www.physics.queensu.ca/$\sim$wade/mimes/}}.

The rotational frequency was deduced from time series of spectro-polarimetric
data or from rotational splitting of oscillation frequencies. It is noteworthy
that, whenever values from these two independent methods are available for a
star, they are in excellent agreement \citep[see][for a recent
  example]{Briquet2013}. Whenever available, we took the $v\sin\,i$-values
  deduced from time-series spectroscopy analysed in such a way that
  line-broadening due to the oscillations was carefully taken into account
  \citep[e.g.,][Chapter 6]{AertsDeCat2003,Aerts2010}.  In the absence of
  time-series spectroscopy or spectro-polarimetry, we relied on the
  $v\sin\,i$-values derived from high-resolution single snapshot spectra.

For the values of $\log\,T_{\rm eff}$, $\log\,g$, and the nitrogen
  abundance, various sources are available in the literature, based on
  different methods and analysis codes.  We relied on recent NLTE analyses of
  high-resolution high signal-to-noise spectroscopy, mostly done by
  \citet{Nieva2012,Martins2012b,Martins2012a,Morel2008}, as well as on detailed
  asteroseismic or spectro-polarimetric studies of individual targets which also
  included NLTE line-fitting of the high-precision spectroscopy. For stars not
  covered in this way, we took the most recent high-precision spectroscopic 
data source available.

Table\,\ref{symbols} gives a summary of the observed properties of the 68 sample
stars, with an indication of the level of completeness of the sampling for
  each of the ten variables $X_1,\ldots,X_{10}$,  while Table\,\ref{data}
contains the actual data of all stars we used in our analyses along with all the
data sources. All stars are plotted in a $(\log\,T_{\rm eff},\log\,g)$ 
diagram in Fig.\,\ref{fig1},
  together with some evolutionary tracks starting from the zero-age
  main-sequence. These were computed with the {\tt MESA}
  stellar evolution code \citep{Paxton2011,Paxton2013}, taking the solar
  composition given by \citet{Asplund2009} and ignoring rotational mixing. We
  adopted the mixing-length theory of convection with a mixing-length value of
  1.8 local pressure scale heights. The Schwarzschild criterion for convection
  was used, along with a fully-mixed core overshoot region extending over 0.15
  local pressure scale heights, which is a good average of all seismically
  determined values for OB pulsators \citep{Aerts2014}.  Among the 68 stars are
  17 unevolved spectroscopic binaries whose orbits are known and for which
  the binarity has appropriately been taken into account in the primary's
  parameters listed in Table\,\ref{data}.  These are indicated with triangles in
  Fig.\,\ref{fig1}.

Some comments on Fig.\,\ref{fig1} are warranted.  \citet{Hubrig2006} already
pointed out the high $\log\,g$ value with large uncertainty for the star
HD\,46005, placing it below the ZAMS. On the other side of low $\log\,g$ values,
we see that only one star in the sample is evolved beyond core-hydrogen
burning. This is the rotational variable B5II star HD\,46769 studied from
time-resolved CoRoT space photometry and high-precision spectroscopy
\citep{Aerts2013a}. The additional seemingly evolved object is the double-lined
spectroscopic eclipsing binary V380\,Cyg, for which a core overshoot value of
0.6 local pressure scale heights was reported to bring the model-independent
measured dynamical mass deduced for the primary in agreement with single-star
evolutionary models \citep{Guinan2000}. This binary is the brightest star
observed by the {\it Kepler\/} spacecraft and these high-precision photometric
space data along with an extensive new set of time-resolved high-resolution
spectroscopy led to dynamical masses with a relative precision near 1\% and
revealed the primary to be a low-amplitude rotational variable with mild silicon
spots and additional stochastic low-frequency photometric variability. These
latest data confirmed the earlier findings that a high core-overshoot parameter
is necessary to bring the primary's mass into agreement with single-star models
and that the secondary does not fit the isochrones for the measured equatorial
rotational velocity and metallicity of both stars \citep{Tkachenko2013}.  This
binary indicates that the current stellar models have too limited near-core
mixing.  As can further be seen in Fig.\,\ref{fig1}, all other stars in our
sample cover the main-sequence phase of evolution.

Finally, the missing data due to lack of measurements were assumed to be missing
at random \citep{Rubin1976,Molenberghs2007}, in line with common statistical
practice.  As is often done in astronomy, we consider logarithmic quantities for
the variables $X_4$, $X_6$, $X_7$, $X_8$, $X_9$, and $X_{10}$, given their wide
ranges of possible values among OB stars, where $X_7$ was transformed in such a
way that a null detection for the magnetic field translates into zero value.

\begin{table}
\begin{center}
\caption{\label{symbols}Characteristics of the observed data set.}
\tablecomments{The magnetic field is assumed to be an oblique
dipole and $B_{\rm pol}$ is
  the polar field strength. ``Tr?''  stands for ``Truncated with upper limit''
  or not.  All variables are positive quantities and are thus truncated with
  zero as lower limit.  The last column gives the percentage of observed
    values expressing the level of completeness within the data set, where we
    considered both measured values and measured limits to determine the
    percentage.  } \tabcolsep=2pt
\vspace*{2mm}
\begin{tabular}{ccllccc}
\hline
Variable & 
Quantity &
Physical meaning &
Unit & 
Observed range &
Tr? & \%\,obs.
\\
\hline
$X_1$ & $v\sin i$ & projected rotational velocity & km/s & [0,310] & no &100\%\\
$X_2$ & $f_{\rm rot}$ & rotational frequency & per day &  [0,2.505] & no &94\%\\
$X_3$ & $f_g$ & gravity mode frequency & per day & [0.039,1.157] & no &32\%\\
$X_4$ & $A_g$ & amplitude of $f_g$ & mmag & [0.11,24.90] & yes &96\%\\
$X_5$ & $f_p$ & acoustic mode frequency & per day & [3.258,10.935] & no&32\% \\
$X_6$ & $A_p$ & amplitude of $f_p$ & mmag & [0.05,38.10] & yes&96\% \\
$X_7$ & $\log (1+B_{\rm pol})$ & magnetic field strength & Gauss & [0.00,4.20]
&yes& 96\%\\
$X_8$ & $\log T_{\rm eff}$ & effective temperature & Kelvin & [4.061,4.633]&no
& 100\%\\ 
$X_9$ &$\log g$ & gravity & cm/s$^2$ &  [2.70,4.43] & no & 100\%\\ \hline
$X_{10}$ & $12+\log [N/H]$ & nitrogen abundance & dex & [7.42,8.95] & yes&
59\% \\
\end{tabular}\end{center}
\end{table}


\section{Statistical Methodology and Analysis}

Handling the incomplete and truncated data was done by multiple imputation
combined with acceptance-rejection sampling
\citep{Rubin1987,Schafer1997,Little2002,CarpenterKenward2013}.  These are
established techniques in medical statistics but less so in astrophysics. We
thus explain first why we used that methodology and how we tuned it to our
application before outlining the various steps involved in the analysis.

The great strength of multiple imputation is that it is a principled
  statistical method that takes proper account of the loss of information in the
  incomplete data. As with any statistical technique it does of course rest on
  assumptions. The role of these assumptions becomes more critical with
  increasing amounts of missing information. However, the connection between
  missing information and the amount of missing data is far from
  straightforward. For example, missing information also depends on the type of
  outcome variables, the patterns of missingness, and the dependence structure
  between what is observed and what is missing, and the model fitted. For these
  reasons, we have limited the data set to the variables listed in
  Table\,\ref{symbols}, where the poorest level of completeness occurs for the
  oscillation frequencies ($X_3$ and $X_5$, which each have a coverage of 32\%)
  but we have well-determined upper limits for their amplitudes from the
  Hipparcos data ($X_4$ and $X_6$).  

Because data come from various sources, there is heterogeneity in the error
  with which they are measured. Moreover, it is well known that good error
  propagation for spectroscopic quantities ($X_1$, $X_7$, $X_8$, $X_9$,
  $X_{10}$) is a difficult problem due to the possible dominance of systematic
  uncertainties over statistical errors. These systematic uncertainties result
  from a combination of instrument calibrations, varying atmospheric conditions,
  spectrum normalization uncertainties, and limitations in the theory of
  spectral line predictions.  Given that our analysis is based on multiple
  regression, in which one works conditionally on the values observed for the
  explanatory variables, such heterogeneity is not explicitly accommodated. The
  implication of this is that estimated relationships may be somewhat attenuated
  \citep{Carroll2010}, but the logic of the analysis is not undermined. In
  particular, ignorance of the measurement errors will not create artefacts such
  as non-existing relationships.

\newcommand{\BY}{\mbox{\boldmath $Y$}}
\newcommand{\BX}{\mbox{\boldmath $X$}}
\newcommand{\bfbeta}{\mbox{\boldmath $\beta$}}
\newcommand{\bfell}{\mbox{\boldmath $\ell$}}
\newcommand{\bfu}{\mbox{\boldmath $u$}}
\newcommand{\bftheta}{\mbox{\boldmath $\theta$}}
\newcommand{\BC}{\mbox{\boldmath $C$}}
\newcommand{\by}{\mbox{\boldmath $y$}}
\newcommand{\bx}{\mbox{\boldmath $x$}}
\newcommand{\bc}{\mbox{\boldmath $c$}}
\newcommand{\bmu}{\mbox{\boldmath $\mu$}}

\subsection{\label{method}Incomplete Data: Missingness and Truncation}

Let $\BX_i=(X_{i1},\dots,X_{ip})^T$ be the column vector of $p$
measurements for star $i=1,\dots,n$. The data are commonly organized in a
dataset $\BX=(\BX_1^T,\dots,\BX_n^T)^T$ 
that takes the form of a rectangular matrix
of dimension $n\times p$. When the data are complete, a wide variety of
statistical models can be fitted to them. As an example, consider a regression
model
\begin{equation}
\label{regres1}
X_{i1}=\beta_1+\beta_2X_{i2}+\cdots+\beta_pX_{ip}+\varepsilon_i,
\end{equation}
where $\bfbeta=(\beta_1,\dots,\beta_p)^T$ is a vector of unknown regression
coefficients and the error term $\varepsilon_i$ is assumed to follow a
distribution with mean zero and variance $\sigma^2$. Using conventional
methodology and standard statistical software, the parameters $\bfbeta$ and
$\sigma^2$ can then be estimated, along with measures of precision, i.e.,
confidence
intervals. 

In the current problem, we are faced with two types of incompleteness. First,
some values $X_{ij}$ are entirely missing. Second, for some values $X_{ij}$
bounds are available per individual star, but not the actual value, i.e.,
the information is restricted to $\ell_{ij}\le X_{ij}\le u_{ij}$, with
$\ell_{ij}$ ($u_{ij}$) a star-specific lower bound (upper bound).  Note
  that, if only one bound is available for all stars simultaneously, then we use
  an additional superscript $r$ to indicate that, i.e., either
  $\ell_{ij}=\ell^r_{ij}$ for one lower bound only or $u_{ij}=u^r_{ij}$ for one
  upper bound only, where $[\ell^r_{ij},u^r_{ij}]$ is then the total range of
  the random variable $X_{ij}$. This occurs, e.g., due to the requirement for
  positive frequency values, leading to zero as lower bound for all stars for
  variables $X_2$, $X_3$, and $X_5$. 
Similarly, we required the values for $X_{10}$ to
  be bound by physically meaningful values as explained in the following
  section.  In summary, it is possible to unify missingness and truncation by
simultaneously setting both bounds to their range limits. Even a properly
observed value can be placed within this setting:
$\ell_{ij}=X_{ij}=u_{ij}$. Thus, the vectors $\bfell_i$, and $\bfu_i$ fully
describe the data available on star~$i$.

Modeling such data has received a large amount of attention.  Here, we opted for
so-called multiple imputation 
\citep{Rubin1987,Schafer1997,Little2002,CarpenterKenward2013}
combined with acceptance-rejection
sampling.  Multiple imputation consists of three steps. In the first or
imputation step, the principle is to replace each missing values with $M$ copies
or so-called imputations. These are drawn from the predictive distribution of
what is missing, given what is observed. Because values are drawn multiple
times, rather than filled in once, the phenomenon that incomplete data lead to
reduced statistical information is maintained, in contrast to single
imputation. 
Let us write the model to be fitted symbolically as
\begin{equation}
f(\bx_i|\bftheta),
\label{genmod}
\end{equation}
where $\bx_i$ is the realized value of $\BX_i$ and $\bftheta$ groups 
all model parameters.
The modeler thus obtains $M$ completed
datasets. In the second or modeling step, each of these is analyzed separately,
as if the data were complete. Thus, $M$ estimates of $\bftheta$, are
obtained. We denote these by $\widehat{\bftheta}_m$, with $m=1,\dots,M$. The
same is true for the corresponding measures of precision. Let the estimated
variance-covariance matrix for $\widehat{\bftheta}_m$ be $\widehat{U}_m$. In the
third or analysis step, these $M$ estimates are combined into a single set of
parameter and precision estimates, using Rubin's rules 
\citep{Rubin1987,Schafer1997,Little2002,CarpenterKenward2013}:
\begin{eqnarray}
\widehat{\bftheta}&=&
\frac{1}{M}\sum_{m=1}^M\widehat{\bftheta}_m,\label{rubinrule1}\\
\widehat{U}&=&
\frac{1}{M}\sum_{m=1}^MU_m+\frac{M+1}{M(M-1)}\sum_{m=1}^M(\widehat{\bftheta}_m-\widehat{\bftheta})(\widehat{\bftheta}_m-\widehat{\bftheta})^T\label{rubinrule2}.
\end{eqnarray}
Step 2 requires fitting a model to a complete set of data, and to repeat this
exercise $M$ times; in step 3, Rubin's rules
(\ref{rubinrule1})--(\ref{rubinrule2}) are applied. The most involved step is
the first one. We turn to it next.

\begin{sloppypar}
For our purposes in particular, the constraints make the use of multiple
imputation non standard, in contrast to when the only complication is
missingness. 
Consider for our purposes the general setting where $\BX_i$ is a
vector with three sub-vectors $\BX_i=(\BX_{i1}^T,\BX_{i2}^T,\BX_{i3}^T)^T$,
where the notation implies that we first transpose 
the individual sub-vectors 
from column to row, place them all next to each other, 
and then turn it into a column once again, in such a way that
$\BX_{i1}$ is observed, $\BX_{i2}$ is truncated with conditions
$\BC_{i2}=\bfell_{i2}\le\BX_{i2}\le\bfu_{i2}$, and $\BX_{i3}$ is fully
missing. Sampling is then needed from
\begin{equation}
\label{ingewikkeld}
f(\bx_{i2},\bx_{i3}|\bx_{i1},\bc_{i2},\bftheta)=f(\bx_{i2},\bx_{i3}|\bx_{i1},\bftheta)\cdot
\frac
{1}
{f(\bc_{i2}|\bx_{i1},\bftheta)}.
\end{equation}
By contrast, should $\BX_{i2}$ as well as $\BX_{i3}$ be fully unobserved, then
it would be sufficient to sample from $f(\bx_{i2},\bx_{i3}|\bx_{i1},\bftheta)$,
the first factor of Eq.\,(\ref{ingewikkeld}).

Assuming a multivariate normal distribution for the variables
in the imputation model implies 
that every conditional distribution of one subset
given another is still multivariate normal  \citep[][Appendix
  B]{JohnsonWichern2000,CarpenterKenward2013}. Hence, 
relying on this convenient property, 
the conditional density is also
multivariate normal and takes the following form:
\begin{equation}
\label{gemakkelijk}
f(\bx_{i2},\bx_{i3}|\bx_{i1},\bftheta)=
\phi_{p_2+p_3}
\left(
\bx_{i,23}
\left|
\bmu_{23}+\Sigma_{23,1}\Sigma_{1,1}^{-1}(\bx_{i1}-\bmu_1)
;
\Sigma_{23,23}-\Sigma_{23,1}\Sigma_{1,1}^{-1}\Sigma_{1,23}
\right.
\right),
\end{equation}
\end{sloppypar}
\noindent 
where $\phi_{p_2+p_3}(\cdot ; \cdot)$ is the multivariate normal density with
dimension $p_2+p_3$. In this notation, an 
index `23' indicates selection of the appropriate components of the full
vector or matrix pertaining to the second and third sub-vector combined.  Not
only is this predictive distribution much simpler than Eq.\,(\ref{ingewikkeld}),
expressions of this form are part of standard implementations of multiple
imputation \citep{CarpenterKenward2013}.

The above considerations suggest combining multiple imputation with so-called
acceptance-rejection sampling 
\citep{vonNeumann1951,Gilks1996,RobertCasella2004}. 
Generally, when sampling from
$h(\bx)$ is required, but sampling from $g(\bx)$ is much easier, then one can
proceed by sampling from the latter density, provided that there is a value
$M>1$ such that $h(\bx)/g(\bx)\le M$.  In our case,
$h(\bx_i)=f(\bx_{i2},\bx_{i3}|\bx_{i1},\bc_{i2},\bftheta)$ while
$g(\bx_i)=f(\bx_{i2},\bx_{i3}|\bx_{i1},\bftheta)$, and the ratio between both is
$M=f(\bc_{i2}|\bx_{i1},\bftheta)^{-1}\ge 1$. 

Acceptance-rejection sampling operates in the following way. Draw $\BX$ from
$g(\bx)$ and $U\sim U(0,1)$, a uniform variable on the unit interval. Then,
accept the draw if $U\le h(\bx)/[Mg(\bx)]$ and reject it otherwise. Given the
form of 
Eq.\,(\ref{ingewikkeld}), $h(\bx)/[Mg(\bx)]=1$ when $\BC_{i2}$ is satisfied
and 0 otherwise. Hence, thanks to the special situation posed by truncation,
every draw that satisfies the constraint is always accepted, otherwise, it is
always rejected.

Thus, in practice, multiple imputations are drawn from a multivariate
distribution
assuming that the missing values and the truncated values are all
missing, but draws are accepted only if they satisfy the constraints $\BC_{i2}$.
The draws that are accepted then form the appropriate predictive
distribution.
Acceptance-rejection sampling can be very inefficient when $g(\cdot)$ and
  $h(\cdot)$ are very different, resulting in small proportions of acceptable
  draws. To improve the efficiency of our imputation procedure, we have
  additionally used transformations that automatically respect range
  restrictions whenever such restrictions are applicable to a variable for all
  stars simultaneously,
  rather than to values for some particular stars only. For example, an
  additional square root transformation was adopted to ensure non-negativity.

\subsection{\label{pca}Principal Components} 

Principal component analysis \citep{Krzanowski1988,JohnsonWichern2000},
abbreviated as PCA, is a classical exploratory method, based on rotating outcome
vectors $\BY_i=L\BX_i$ with the requirement that the components of the
transformed vectors are uncorrelated, that the first one has maximal variance,
the second one maximal variance given the first, etc.  Ordinarily, PCA is
conducted on standardized, hence unitless, input variables $\BX_i$. Technically,
the transformation matrix $L$ is the set of eigenvectors of the correlation
matrix of the input variables $\BX_i$. The variances of the transformed
variables are found as the eigenvalues of the said correlation matrix. For
example, the first principal component
\begin{equation}
Y_{i1}=\ell_{11}X_{i1}+\dots+\ell_{1p}X_{ip}
\end{equation}
is determined by the first eigenvector $\bfell_1=(\ell_{11},\dots,\ell_{1p})^T$
and its variance is $\lambda_1$, the leading eigenvalue. The coefficients of
$\bfell_1$ indicate the importance with which the observed variables $X_{ij}$
contribute to the first principal component, with the same logic applying to the
other principal components. Further, $\lambda_1/p$ is the fraction of the total
variance in the original, standardized variables that is captured by the first
principal components.

\subsection{Application to the Selected Sample}

For our application, variables $X_1$, $X_8$, and $X_9$ 
($v\sin\,i$, $\log\,T_{\rm eff}$, and $\log\,g$)
are observed for all 68
stars, while all others have missing values (see Table\,\ref{symbols}). 
By definition, all variables must
be positive and are thus truncated at zero as their lower limit, but we use the
measured lower limits for $X_7$ (the magnetic field strength)
when available.  Furthermore, there is
truncation as an upper limit on variables $X_4$, $X_6$, and $X_{10}$
(the oscillation mode amplitudes and the nitrogen abundance). For
$X_{10}$, we required physically meaningful results in that imputed values must
be contained in $[6.8,9.0]$ following results achieved for the Milky Way and the
Magellanic Clouds \citep{Trundle2007,Rivero2012}.

We fit a linear regression model to the nitrogen abundance ($X_{10}$), with
$X_1$, $X_2$, $X_3$, $X_4$, $X_5$, $X_6$, $X_7$, $X_8$, and $X_9$ as potential
predictors.  
Although obviously from a physical perspective a linear model 
does not necessarily reflect the underlying theory,
it does have specific advantages that are appropriate in
this setting. First, a linear regression does not rest on a priori theory and
hence is neutral with respect 
to the specific relationships uncovered. Second, it is not the
aim to establish an entire physical theory for the processes being modelled, but
rather to bring to light important relationships that would be hard to unravel
when considering only one variate at a time and that can then be explored
further.

Twenty imputations were drawn, using the Monte Carlo Markov Chain
method \citep{Neter1996}. 
The advantage of this method is that it can easily handle
missingness of a non-monotone type, i.e., where arbitrary patterns of
missingness occur among the stars. It is well-known that small numbers of
imputations typically produce valid 
results, and \citet{Little2002} even quote five as sufficient
choice. This is relevant for very large data bases, where drawing imputations is
computationally costly, while here the number of imputations is not really of
concern, given that there are only 68 stars. On the other hand,
acceptance-rejection sampling with bounds applying to the variables,
necessitates drawing large numbers of imputations. For some stars, 5,000
imputations were needed to obtain the requested number of 20 non-rejected
imputations. Even then, for 4 of the 68 stars, no valid draws could be
found. These stars are indicated in Table\,\ref{data}.  
A sensitivity analysis was conducted to
address this issue (see Section\,\ref{sensitivity}), as is common practice
  in statistical modeling.

A regression model was then fitted to the data by first fitting it to all 20
imputations and then combining the results using the rules outlined in
Section\,\ref{method}.  To
select a well-fitting, parsimonious model, two routes were followed.  In the
first, backward selection \citep{Neter1996}, all nine predictor variables were
included. Then, the least significant one was deleted, the model re-fitted, and
the process repeated until only significant predictors remained.  Significance
is defined at the conventional five percent level (indicated as $p<0.05$). It
turns out that only two predictors ($X_5$ and $X_8$, i.e., $f_p$ and
$\log\,T_{\rm eff}$) are significant (top part
of Table\,\ref{result}).  
In the second, forward selection starts from including only the
most significant predictor at first, which is $X_8$ as can be seen from the
top left-hand columns in Table\,\ref{result}. 
Thereafter, the model is re-fitted with $X_8$
included by default, paired with each one of the remaining predictors in
turn. The most significant of these is again $X_5$. The process was iterated but
no further significant predictors could be added. While generally backward and
forward selection can lead to very different models, here the same two
predictors for the nitrogen abundance ($X_{10}$) are selected with both
approaches, i.e., the effective temperature ($X_8$) and the frequency of the
dominant acoustic oscillation mode ($X_5$).  The top part of Table\,\ref{result}
describes
the results of the model selection processes.
\begin{figure}[t!]
\begin{center}
\rotatebox{270}{\resizebox{12cm}{!}{\includegraphics{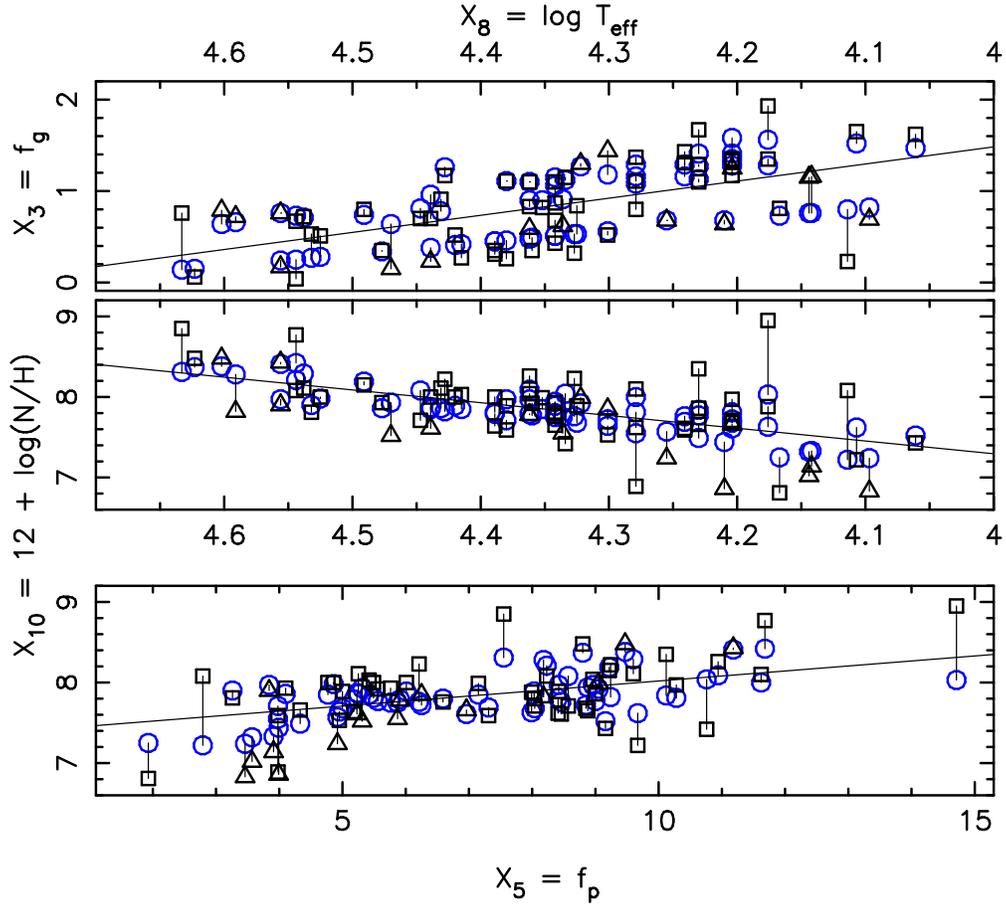}}}
\end{center}
\caption[]{Observed or averaged 
imputed values (squares for single stars, triangles for
  spectroscopic binaries) connected by lines to the values predicted by the
  joint models described in Table\,\ref{result} (circles, blue in the online
  version of the paper) for the gravity-mode frequency ($X_3$, upper
    panel) and for the nitrogen abundance ($X_{10}$, lower two panels).  The
    full lines represent the univariate model fits listed in the left part of
    Table\,\ref{result}.}
\label{fig2}
\end{figure}
\begin{figure}[h!]
\begin{center}
\rotatebox{270}{\resizebox{12cm}{!}{\includegraphics{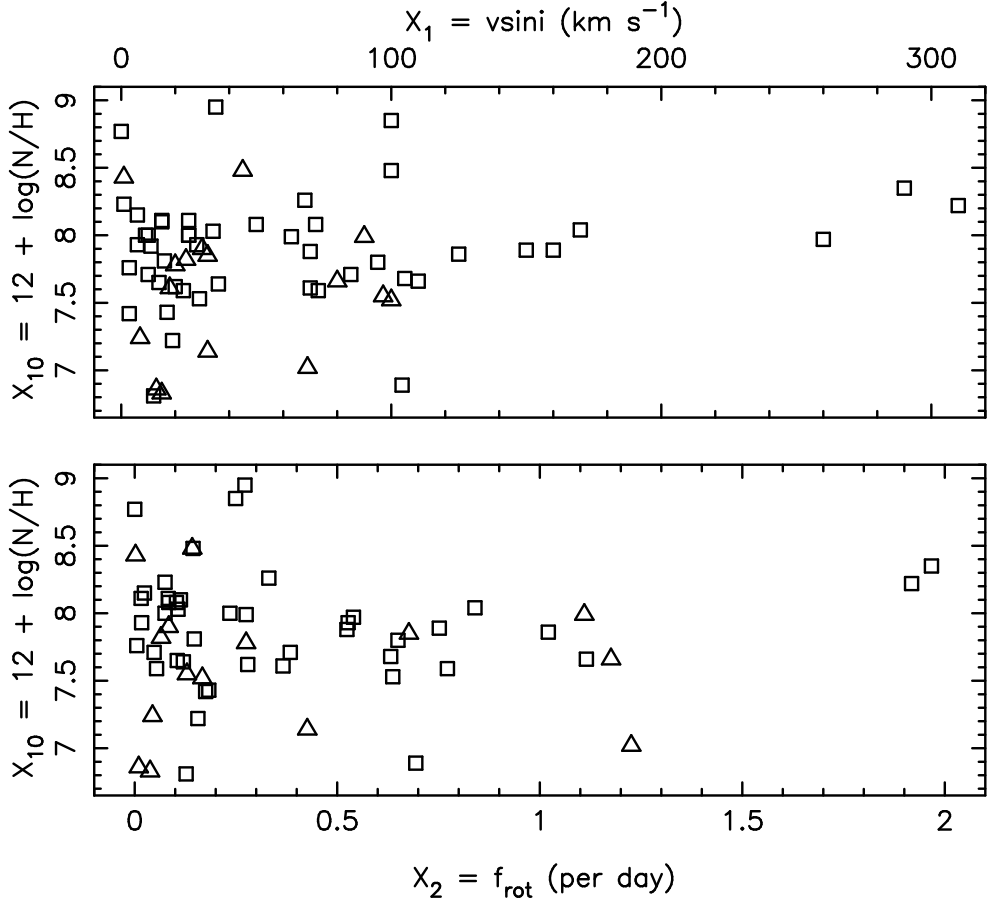}}}
\end{center}
\caption[]{Observed or averaged 
imputed values (squares for single stars, triangles for
  spectroscopic binaries) for the nitrogen abundance ($X_{10}$) as a
    function of $v\sin\,i$ and the rotation frequency ($X_1$ and $X_2$).}
\label{fig3}
\end{figure}

Returning to the left-hand side of Table\,\ref{result}, 
these are the coefficients (with
standard errors in brackets), and significance levels when the two
predictors are considered one at a time. Qualitatively, $X_5$, i.e., the
frequency of the dominant acoustic mode, is a significant predictor on its own
for the nitrogen abundance ($X_{10}$). The fraction of the variance
  explained by the model (denoted as $R^2$) for
this single predictor ranges up to 36\% for the nitrogen
abundance. The effective temperature ($X_8$) explains up to 37\% of the
variance of $X_{10}$ and is also a significant predictor on its own. 

The lower two panels of Figure\,\ref{fig2} present an illustration of the
quality of the joint model. The observed or averaged imputed values of $X_{10}$
(squares for single stars, triangles for spectroscopic binaries) are connected
with those predicted by the joint model for $X_{10}$ given in the upper part of
Table\,\ref{result} (circles), as a function of $X_8$ (middle panel) and of
$X_5$ (lower panel).  Although the joint model is satisfactory, the
corresponding fraction of the variance explained by the model across the
20 imputations is at best 54\%.

The results of our statistical modeling lead to the following hypotheses.
  The acoustic mode frequencies are a measure of the mean density of the star
  \citep[e.g.,][]{Aerts2010}.  Following the mass-radius relation that holds
  during core-hydrogen burning, the acoustic frequencies decrease as the mass of
  the star increases.  On the other hand, the effective temperature increases as
  the mass increases. From the coefficients of the joint model for the nitrogen
  abundance in Table\,\ref{result} and the range of values for $X_5$ and $X_8$
  as given in Table\,\ref{symbols}, we find that the effect of the temperature
  is dominant over the one of the acoustic mode.  Thus, among the sample of OB
  stars considered here, the higher-mass O stars have more nitrogen enrichment
  than the lower-mass B stars, as also found in the study 
by \citet{Rivero2012}, 
but the presence of acoustic modes further seems
  to increase the nitrogen abundance and this increase due to the oscillations
  gets larger as we move from the early O stars to the late B stars.

To investigate inter-relationships among the variables as well as the
  variance in the data set as a whole, without focusing solely on the nitrogen
abundance, we performed a principal component analysis
\citep{Krzanowski1988,JohnsonWichern2000}, as discussed in Section\,\ref{pca}.
This revealed evidence of two relationships: a first one between the dominant
gravity-mode frequency ($X_3$), the magnetic field strength ($X_7$), the
effective temperature ($X_8$), and the rotational variables ($X_1$ and $X_2$)
explaining 29\% of the variance in the data set and a second one between
the nitrogen abundance ($X_{10}$), some of the mode properties ($X_4$, $X_5$)
and the effective temperature ($X_8$) explaining 19\% of the variance.  It
is reassuring that the joint model for the nitrogen abundance is recovered in
this way from the second principal component. Following the first principal
component, a joint model analysis was repeated for variable $X_3$ ($f_g$)
and led to the
results in bottom part of Table\,\ref{result}: the lower the effective
temperature and the stronger the magnetic field, the higher the gravity-mode
frequency, while the rotational frequency and projected rotational velocity were
found to be insignificant as predictors for $X_3$.

The upper panel of Figure\,\ref{fig2} is an illustration of the quality of
  this joint model for $X_3$.  Again, the observed or averaged imputed values of
  $X_3$ are connected to those predicted by the joint model given in the lower
  part of Table\,\ref{result} (circles), as a function of $X_8$.  In this case,
  the joint model across the 20 imputations explains up to 65\% of the variance
  in the frequency of the dominant gravity mode. The increase of the
  gravity-mode frequencies with decrease of the effective temperature was
  already reported by \citet[][their Fig.\,18]{DeCatAerts2002} from their much
  smaller sample of slowly pulsating B stars, which covers only a very narrow
  range in effective temperature compared to the sample we composed here.  In
  addition, we find a correlation between the gravity-mode frequency and the
  magnetic field strength. As far as we are aware, such a connection has not
  been reported before from observational diagnostics,
although tight relationships
between low-frequency gravity modes and the magnetic field
  strength have been presented in theoretical work \citep[e.g.,][]{Hermans1990}.
Of course, we must keep in mind that the spectro-polarimetric measurements
  upon which we relied in this work reveal only the topology and strength of the
  magnetic field in the outer layers of the star while
the internal properties may
  be quite different.  Moreover, 
the relationship we find results directly from the imputed values of the
  magnetic field strength and gravity-mode frequency. Indeed, it cannot be
  revealed without imputation, because very few stars with both gravity-mode
  oscillations and a positive magnetic field strength have been detected.
  In our sample, e.g., 20 stars have both magnetic field measurements and
  detected gravity modes, but for 18 of those a null detection for the magnetic
  field strength occurs. Hence few gravity-mode oscillators have a magnetic
  field, but the imputed values for $X_7$ for all stars suggest a correlation
  between $X_3$ and $X_7$ which remains to be checked in the future from new
  magnetic gravity-mode pulsators yet to be discovered.

\subsection{\label{sensitivity}Sensitivity Analysis}

The fact that four stars were removed from the joint analyses out of necessity
needs to be addressed. We performed a so-called sensitivity analysis by also
considering multiple imputation and acceptance-rejection without requiring zero
lower bounds and without transforming the data to their square-root value.  This
allowed all 68 stars to be included and led to very similar results, although
the frequency of rotation ($X_2$) or of the gravity mode ($X_3$) for some of
these stars were assigned negative imputed values.
\begin{table}
\begin{center}
\caption{\label{result}Parameter estimates (standard errors), $p$-values, and
  the fraction of the variance 
explained by the model $(R^2)$.}
\tablecomments{\ Left 
hand columns: separate models with a single predictor; right hand columns:
  joint model with significant predictors simultaneously included. The top part
  concerns models for the nitrogen abundance ($X_{10}$) and the bottom part for
  the frequency of the dominant gravity-mode oscillation ($X_3$). (Pred.:
  predictor variable).}  \tabcolsep=4pt
\begin{tabular}{lcrrccrrc}
\hline
&&\multicolumn{3}{c}{Separate models}&&\multicolumn{3}{c}{Joint model}\\
\cline{3-5}\cline{7-9}
\multicolumn{1}{c}{Pred.}&&
\multicolumn{1}{c}{Estimate~(s.e.)}&
\multicolumn{1}{c}{$p$}&
\multicolumn{1}{c}{$R^2$ range}&&
\multicolumn{1}{c}{Estimate~(s.e.)}&
\multicolumn{1}{c}{$p$}&
\multicolumn{1}{c}{$R^2$ range}\\
\hline\hline
\multicolumn{9}{c}{Models for $X_{10}$}\\
\hline
intercept&&     ---           & ---     &---&& 0.6744(1.9285)&---&\\ 
\cline{3-5} 
$X_5$    && 0.0622(0.0225)&0.0068&[0.05;0.36]&& 0.0595(0.0217)&   0.0079&\\
\cline{3-5} 
$X_8$    && 1.5848(0.5009)&0.0018&[0.08;0.37]&& 1.5515(0.4428)&   0.0006\\
&&&&&&&&[0.21;0.54]\\
\hline\hline
\multicolumn{9}{c}{Models for $X_3$}\\
\hline
intercept&&     ---           & ---     &---&& 6.0323(2.2180)&---&\\ 
\cline{3-5} 
$X_7$    && 0.1801(0.0412)&$<$0.0001&[0.26;0.56]&& 0.1546(0.0386)&   0.0001&\\
\cline{3-5} 
$X_8$    &&-1.8711(0.5888)&0.0018&[0.11;0.31]&&-1.2723(0.5087)&   0.0136\\
&&&&&&&&[0.34;0.65]\\
\end{tabular}
\end{center}
\end{table}

\section{Discussion}

The theory of single-star 
evolution relying on rotational mixing predicts that
  the surface nitrogen abundance of a star ($X_{10}$) should strongly depend on
  its equatorial rotation velocity. Hence, recent observational studies aimed to
  test the theoretical predictions mainly considered the observed projected
  rotational velocity ($X_1$) and corrected it by assuming equal probability of
  all possible inclination angles to take away the dependence on the unknown
  factor $\sin\,i$
  \citep[e.g.,][]{Hunter2008,Brott2011,Rivero2012,Bouret2012,Bouret2013}.  Those
  studies led to the conclusion that the theoretical models fail to
  explain the observed values of the nitrogen abundance.  Here, we come to a
similar conclusion, by relying on {\it a multitude of observed stellar
  properties of 64 galactic OB stars}. In particular, our sample contains a
  majority of stars with $v\sin\,i$ below 100\,km\,s$^{-1}$ while their
  nitrogen abundance ranges from about 7 to almost 9\,dex. 
We deduce that neither
the projected rotational velocity ($X_1$) nor the rotational frequency ($X_2$)
has predictive power for the measured nitrogen abundance.  This is illustrated
graphically in Fig.\,\ref{fig3}.  Comparison of the middle and lower panels of
Fig.\,\ref{fig2} and both panels of Fig.\,\ref{fig3} visually shows that the
rotational parameters are indeed less suitable as predictors of the nitrogen
abundance compared to the effective temperature and the dominant acoustic
oscillation mode frequency.  

In fact, our results imply that none of the nine variables considered here is by
itself able to predict more than a third of the variance in the nitrogen
abundance.  The joint bivariate model developed for the nitrogen abundance
  listed in Table\,\ref{result}, i.e.,
\begin{equation} 
12+\log [N/H] = 0.6744(1.9285) + 0.0595(0.0217)\cdot f_p + 
1.5515(0.4428)\cdot \log\,T_{\rm eff}
\label{regressie}
\end{equation}
offers an appropriate {\it yet still incomplete\/} comparison between theory
and observations, for the range of stellar parameters in the considered galactic
sample.

We come to the conclusion that mixing must occur already early-on in the
core-hydrogen burning phase for a considerable fraction of massive stars and
that it cannot be due to rotational mixing alone.  The same conclusion was
  reached by an independent study of LMC O stars by \citet{Rivero2012}.  Other
dynamical phenomena causing mixing must thus be considered. From the
  statistical modeling based on the observed quantities treated in this work,
  heat-driven oscillation mode frequencies are a more suitable predictor for
mixing compared to the magnetic field strength and the rotation in the sample of
stars we studied here. Hence we suggest to include the oscillation properties
into future evaluations of theoretical model predictions, e.g., by means of
Eq.(\ref{regressie}), rather than or in addition to the rotational properties.

Whether our conclusions hold for evolved massive stars and/or for metal-poor
stars cannot be checked at present, because the asteroseismic data needed for
such a study is lacking. Unfortunately, relatively few massive stars have been
monitored sufficiently long in high-precision space photometry because such
objects hardly occurred in the fields-of-view of the CoRoT and {\it Kepler\/}
space missions, given that these satellites focused on exoplanet hunting around
faint cool stars in rather crowded fields, for which the presence of bright
massive stars is a nuissance.  It will thus require large dedicated and
coordinated ground-based efforts to remedy this situation, because long-term
monitoring is necessary to reveal and unravel the oscillation properties of
evolved OB stars.  The potential of such studies from space photometry is clear
and it could be tackled with the {\it Kepler2.0\/} mission concept \citep[see,
  e.g.,][]{Aerts2013b} and with the future PLATO2.0 mission \citep{Rauer2013}.

Apart from the observed acoustic and gravity modes excited by a heat-mechanism
in the stellar envelope upon which we relied in this work, large-scale
low-frequency internal gravity waves excited by the convective core, as in the
studies by \citet{Browning2004,Neiner2012,Rogers2013,Mathis2013}, may contribute
considerably to the surface nitrogen abundance if these waves are able to
propagate through the acoustic mode cavity and to transport chemical species in
addition to angular momentum.  So far, theoretical or simulation studies focused
on the angular momentum transport of such internal gravity waves but it should
be well possible to extend those studies to the chemical transport they may
induce.  The frequency spectrum of internal gravity waves computed from 2D
simulations for a 3\,M$_\odot$ star by \citet{Rogers2013} coincides with the
frequency range of running waves expected for OB stars and leads to the
conclusion that the convective penetration depth {\it decreases\/} as a function
of the rotation frequency. This result is in disagreement with the one based on
3D simulations for a 2\,M$_\odot$ star by \citet{Browning2004}, who find the
penetration depth to be {\it increasing\/} with rotation frequency
\citep[see][for a discussion of the origin of this discrepancy]{Mathis2013bis}.
Clearly, a better understanding of such type of simulations for stars with a
convective core is needed.  Nevertheless, both simulation studies do find a
range of the penetration depth and of core overshooting in agreement with the
findings from asteroseismology based on standing waves \citep{Aerts2014}.
Moreover, these simulation studies show that {\it internal gravity waves 
transport angular momentum, thereby decreasing the level of differential
  rotation and hence counteract the effect of rotational mixing as the stellar
  evolution progresses}.  We therefore consider the simultaneous act of internal
gravity waves and differential rotation to be the best explanation for the
measured surface nitrogen abundances in massive stars, despite the fact that the
chemical transport that such waves may induce remains to be computed.  At
present, direct observational velocity diagnostics resulting from internal
gravity waves are at best speculative in terms of macroturbulent spectral line
broadening \citep[e.g.,][]{Aerts2009,SimonDiaz2010}.  As soon as velocity
signatures due to internal gravity waves can be established with certainty, they
can be added as variables for joint statistical modeling, in addition to the
signatures of heat-driven acoustic and gravity mode oscillations that were used
in the current work.

\acknowledgments The authors are grateful to the MiMeS collaboration for having
provided magnetic field null detections prior to publication.  They also
acknowledge the numerous constructive comments by the referee, Dr.\ Joachim
Puls, which helped to improve the presentation of this work appreciably.  The
research leading to these results has received funding from the European
Research Council under the European Community's Seventh Framework Programme
(FP7/2007--2013)/ERC grant agreement n$^\circ$227224 (PROSPERITY) and from IAP
research Network P7/06 of the Belgian Government (Belgian Science Policy). Conny
Aerts is grateful to the Lorentz Center in Leiden, the Netherlands, 
for providing
a stimulating research environment by means of its workshop programme; the idea
for this study found its origin during the workshop ``Steps Towards a New
Generation of Stellar Models''. \\[0.2cm]

\bibliographystyle{aa}
\bibliography{ms}

\begin{thebibliography}{133}
\expandafter\ifx\csname natexlab\endcsname\relax\def\natexlab#1{#1}\fi

\bibitem[{{Aerts}(2014)}]{Aerts2014}
{Aerts}, C. 2014, in ``Setting a New Standard in the Analysis of Binary
  Stars'', EAS Publication Series, in press (arXiv1311.6242)

\bibitem[{{Aerts} {et~al.}(2011){Aerts}, {Briquet}, {Degroote}, {Thoul}, \&
  {van Hoolst}}]{Aerts2011}
{Aerts}, C., {Briquet}, M., {Degroote}, P., {Thoul}, A., \& {van Hoolst}, T.
  2011, \aap, 534, A98 (75)

\bibitem[{{Aerts} {et~al.}(2010){Aerts}, {Christensen-Dalsgaard}, \&
  {Kurtz}}]{Aerts2010}
{Aerts}, C., {Christensen-Dalsgaard}, J., \& {Kurtz}, D.~W. 2010,
  {Asteroseismology, Astronomy and Astrophsyics Library, Springer Berlin
  Heidelberg}

\bibitem[{{Aerts} \& {De Cat}(2003)}]{AertsDeCat2003}
{Aerts}, C. \& {De Cat}, P. 2003, \ssr, 105, 453

\bibitem[{{Aerts} {et~al.}(1998){Aerts}, {De Cat}, {Cuypers}, {Becker},
  {Mathias}, {De Mey}, {Gillet}, \& {Waelkens}}]{Aerts1998}
{Aerts}, C., {De Cat}, P., {Cuypers}, J., {et~al.} 1998, \aap, 329, 137 (55)

\bibitem[{{Aerts} {et~al.}(2006){Aerts}, {De Cat}, {Kuschnig}, {Matthews},
  {Guenther}, {Moffat}, {Rucinski}, {Sasselov}, {Walker}, \&
  {Weiss}}]{Aerts2006}
{Aerts}, C., {De Cat}, P., {Kuschnig}, R., {et~al.} 2006, \apjl, 642, L165 (7)

\bibitem[{{Aerts} {et~al.}(1999){Aerts}, {De Cat}, {Peeters}, {Decin}, {De
  Ridder}, {Kolenberg}, {Meeus}, {Van Winckel}, {Cuypers}, \&
  {Waelkens}}]{Aerts1999}
{Aerts}, C., {De Cat}, P., {Peeters}, E., {et~al.} 1999, \aap, 343, 872 (57)

\bibitem[{{Aerts} {et~al.}(2003{\natexlab{a}}){Aerts}, {Lehmann}, {Briquet},
  {Scuflaire}, {Dupret}, {De Ridder}, \& {Thoul}}]{Aerts2003b}
{Aerts}, C., {Lehmann}, H., {Briquet}, M., {et~al.} 2003{\natexlab{a}}, \aap,
  399, 639 (89)

\bibitem[{{Aerts} {et~al.}(2009){Aerts}, {Puls}, {Godart}, \&
  {Dupret}}]{Aerts2009}
{Aerts}, C., {Puls}, J., {Godart}, M., \& {Dupret}, M.-A. 2009, \aap, 508, 409

\bibitem[{{Aerts} {et~al.}(2013{\natexlab{a}}){Aerts}, {Sim{\'o}n-D{\'{\i}}az},
  {Catala}, {Neiner}, {Briquet}, {Castro}, {Schmid}, {Scardia}, {Rainer},
  {Poretti}, {P{\'a}pics}, {Degroote}, {Bloemen}, {{\O}stensen}, {Auvergne},
  {Baglin}, {Baudin}, {Michel}, \& {Samadi}}]{Aerts2013a}
{Aerts}, C., {Sim{\'o}n-D{\'{\i}}az}, S., {Catala}, C., {et~al.}
  2013{\natexlab{a}}, \aap, 557, A114 (40)

\bibitem[{{Aerts} {et~al.}(2003{\natexlab{b}}){Aerts}, {Thoul},
  {Daszy{\'n}ska}, {Scuflaire}, {Waelkens}, {Dupret}, {Niemczura}, \&
  {Noels}}]{Aerts2003a}
{Aerts}, C., {Thoul}, A., {Daszy{\'n}ska}, J., {et~al.} 2003{\natexlab{b}},
  Science, 300, 1926 (63)

\bibitem[{{Aerts} {et~al.}(2004){Aerts}, {Waelkens},
  {Daszy{\'n}ska-Daszkiewicz}, {Dupret}, {Thoul}, {Scuflaire}, {Uytterhoeven},
  {Niemczura}, \& {Noels}}]{Aerts2004}
{Aerts}, C., {Waelkens}, C., {Daszy{\'n}ska-Daszkiewicz}, J., {et~al.} 2004,
  \aap, 415, 241 (62)

\bibitem[{{Aerts} {et~al.}(2013{\natexlab{b}}){Aerts}, {Zwintz},
  {Marcos-Arenal}, {Moravveji}, {Degroote}, {Papics}, {Tkachenko}, {De Ridder},
  {Briquet}, {Thoul}, {Saesen}, {Mowlavi}, {Barblan}, {Neiner}, {Pavlovski}, \&
  {Guzik}}]{Aerts2013b}
{Aerts}, C., {Zwintz}, K., {Marcos-Arenal}, P., {et~al.} 2013{\natexlab{b}},
  arXiv1309.3042

\bibitem[{{Alecian} {et~al.}(2011){Alecian}, {Kochukhov}, {Neiner}, {Wade}, {de
  Batz}, {Henrichs}, {Grunhut}, {Bouret}, {Briquet}, {Gagne}, {Naze}, {Oksala},
  {Rivinius}, {Townsend}, {Walborn}, {Weiss}, \& {Mimes
  Collaboration}}]{Alecian2011}
{Alecian}, E., {Kochukhov}, O., {Neiner}, C., {et~al.} 2011, \aap, 536, L6 (54)

\bibitem[{{Asplund} {et~al.}(2009){Asplund}, {Grevesse}, {Sauval}, \&
  {Scott}}]{Asplund2009}
{Asplund}, M., {Grevesse}, N., {Sauval}, A.~J., \& {Scott}, P. 2009, \araa, 47,
  481

\bibitem[{{Blomme} {et~al.}(2011){Blomme}, {Mahy}, {Catala}, {Cuypers},
  {Gosset}, {Godart}, {Montalban}, {Ventura}, {Rauw}, {Morel}, {Degroote},
  {Aerts}, {Noels}, {Michel}, {Baudin}, {Baglin}, {Auvergne}, \&
  {Samadi}}]{Blomme2011}
{Blomme}, R., {Mahy}, L., {Catala}, C., {et~al.} 2011, \aap, 533, A4 (35)

\bibitem[{{Bohlender} \& {Landstreet}(1990)}]{Bohlender1990}
{Bohlender}, D.~A. \& {Landstreet}, J.~D. 1990, \apj, 358, 274 (50)

\bibitem[{{Bohlender} \& {Monin}(2011)}]{Bohlender2011}
{Bohlender}, D.~A. \& {Monin}, D. 2011, \aj, 141, 169 (74)

\bibitem[{{Bohlender} {et~al.}(2010){Bohlender}, {Rice}, \&
  {Hechler}}]{Bohlender2010}
{Bohlender}, D.~A., {Rice}, J.~B., \& {Hechler}, P. 2010, \aap, 520, A44 (59)

\bibitem[{{Bolton} {et~al.}(1998){Bolton}, {Harmanec}, {Lyons}, {Odell}, \&
  {Pyper}}]{Bolton1998}
{Bolton}, C.~T., {Harmanec}, P., {Lyons}, R.~W., {Odell}, A.~P., \& {Pyper},
  D.~M. 1998, \aap, 337, 183 (19)

\bibitem[{{Bouret} {et~al.}(2012){Bouret}, {Hillier}, {Lanz}, \&
  {Fullerton}}]{Bouret2012}
{Bouret}, J.-C., {Hillier}, D.~J., {Lanz}, T., \& {Fullerton}, A.~W. 2012,
  \aap, 544, A67

\bibitem[{{Bouret} {et~al.}(2013){Bouret}, {Lanz}, {Martins}, {Marcolino},
  {Hillier}, {Depagne}, \& {Hubeny}}]{Bouret2013}
{Bouret}, J.-C., {Lanz}, T., {Martins}, F., {et~al.} 2013, \aap, 555, A1

\bibitem[{{Briquet} {et~al.}(2011){Briquet}, {Aerts}, {Baglin}, {Nieva},
  {Degroote}, {Przybilla}, {Noels}, {Schiller}, {Vu{\v c}kovi{\'c}}, {Oreiro},
  {Smolders}, {Auvergne}, {Baudin}, {Catala}, {Michel}, \&
  {Samadi}}]{Briquet2011}
{Briquet}, M., {Aerts}, C., {Baglin}, A., {et~al.} 2011, \aap, 527, A112 (36)

\bibitem[{{Briquet} {et~al.}(2004){Briquet}, {Aerts}, {L{\"u}ftinger}, {De
  Cat}, {Piskunov}, \& {Scuflaire}}]{Briquet2004}
{Briquet}, M., {Aerts}, C., {L{\"u}ftinger}, T., {et~al.} 2004, \aap, 413, 273
  (46)

\bibitem[{{Briquet} {et~al.}(2007{\natexlab{a}}){Briquet}, {Hubrig},
  {Sch{\"o}ller}, \& {De Cat}}]{Briquet2007a}
{Briquet}, M., {Hubrig}, S., {Sch{\"o}ller}, M., \& {De Cat}, P.
  2007{\natexlab{a}}, Astronomische Nachrichten, 328, 41 (47)

\bibitem[{{Briquet} {et~al.}(2005){Briquet}, {Lefever}, {Uytterhoeven}, \&
  {Aerts}}]{Briquet2005}
{Briquet}, M., {Lefever}, K., {Uytterhoeven}, K., \& {Aerts}, C. 2005, \mnras,
  362, 619 (67)

\bibitem[{{Briquet} {et~al.}(2007{\natexlab{b}}){Briquet}, {Morel}, {Thoul},
  {Scuflaire}, {Miglio}, {Montalb{\'a}n}, {Dupret}, \& {Aerts}}]{Briquet2007b}
{Briquet}, M., {Morel}, T., {Thoul}, A., {et~al.} 2007{\natexlab{b}}, \mnras,
  381, 1482 (68)

\bibitem[{{Briquet} {et~al.}(2012){Briquet}, {Neiner}, {Aerts}, {Morel},
  {Mathis}, {Reese}, {Lehmann}, {Costero}, {Echevarria}, {Handler}, {Kambe},
  {Hirata}, {Masuda}, {Wright}, {Yang}, {Pintado}, {Mkrtichian}, {Lee}, {Han},
  {Bruch}, {De Cat}, {Uytterhoeven}, {Lefever}, {Vanautgaerden}, {de Batz},
  {Fr{\'e}mat}, {Henrichs}, {Geers}, {Martayan}, {Hubert}, {Thizy}, \&
  {Tijani}}]{Briquet2012}
{Briquet}, M., {Neiner}, C., {Aerts}, C., {et~al.} 2012, \mnras, 427, 483 (70)

\bibitem[{{Briquet} {et~al.}(2013){Briquet}, {Neiner}, {Leroy}, \&
  {P{\'a}pics}}]{Briquet2013}
{Briquet}, M., {Neiner}, C., {Leroy}, B., \& {P{\'a}pics}, P.~I. 2013, \aap,
  557, L16 (27)

\bibitem[{{Briquet} {et~al.}(2009){Briquet}, {Uytterhoeven}, {Morel}, {Aerts},
  {De Cat}, {Mathias}, {Lefever}, {Miglio}, {Poretti}, {Mart{\'{\i}}n-Ruiz},
  {Papar{\'o}}, {Rainer}, {Carrier}, {Guti{\'e}rrez-Soto}, {Valtier}, {Benk{\H
  o}}, {Bogn{\'a}r}, {Niemczura}, {Amado}, {Su{\'a}rez}, {Moya},
  {Rodr{\'{\i}}guez-L{\'o}pez}, \& {Garrido}}]{Briquet2009}
{Briquet}, M., {Uytterhoeven}, K., {Morel}, T., {et~al.} 2009, \aap, 506, 269
  (76)

\bibitem[{{Brott} {et~al.}(2011){Brott}, {Evans}, {Hunter}, {de Koter},
  {Langer}, {Dufton}, {Cantiello}, {Trundle}, {Lennon}, {de Mink}, {Yoon}, \&
  {Anders}}]{Brott2011}
{Brott}, I., {Evans}, C.~J., {Hunter}, I., {et~al.} 2011, \aap, 530, A116

\bibitem[{{Browning} {et~al.}(2004){Browning}, {Brun}, \&
  {Toomre}}]{Browning2004}
{Browning}, M.~K., {Brun}, A.~S., \& {Toomre}, J. 2004, \apj, 601, 512

\bibitem[{{Bychkov} {et~al.}(2005){Bychkov}, {Bychkova}, \&
  {Madej}}]{Bychkov2005}
{Bychkov}, V.~D., {Bychkova}, L.~V., \& {Madej}, J. 2005, \aap, 430, 1143 (15)

\bibitem[{{Carpenter} \& {Kenward}(2013)}]{CarpenterKenward2013}
{Carpenter}, J.~R. \& {Kenward}, M.~G. 2013, {Multiple Imputation and its
  Application, John Wiley \& Sons, Chichester}

\bibitem[{{Carroll} {et~al.}(2010){Carroll}, {Ruppert}, {Stefanski}, \&
  {Crainiceanu}}]{Carroll2010}
{Carroll}, R.~J., {Ruppert}, D., {Stefanski}, L.~A., \& {Crainiceanu}, C.~M.
  2010, Measurement Error in Nonlinear Models (2nd edition), CRC/Chapman \&
  Hall

\bibitem[{{Chapellier} {et~al.}(1995){Chapellier}, {Le Contel}, {Le Contel},
  {Sareyan}, \& {Valtier}}]{Chapellier1995}
{Chapellier}, E., {Le Contel}, J.~M., {Le Contel}, D., {Sareyan}, J.~P., \&
  {Valtier}, J.~C. 1995, \aap, 304, 406 (91)

\bibitem[{{Cidale} {et~al.}(2007){Cidale}, {Arias}, {Torres}, {Zorec},
  {Fr{\'e}mat}, \& {Cruzado}}]{Cidale2007}
{Cidale}, L.~S., {Arias}, M.~L., {Torres}, A.~F., {et~al.} 2007, \aap, 468, 263
  (64)

\bibitem[{{Cuypers} {et~al.}(2002){Cuypers}, {Aerts}, {Buzasi}, {Catanzarite},
  {Conrow}, \& {Laher}}]{Cuypers2002}
{Cuypers}, J., {Aerts}, C., {Buzasi}, D., {et~al.} 2002, \aap, 392, 599 (56)

\bibitem[{{De Cat} \& {Aerts}(2002)}]{DeCatAerts2002}
{De Cat}, P. \& {Aerts}, C. 2002, \aap, 393, 965 (10)

\bibitem[{{De Cat} {et~al.}(2007){De Cat}, {Briquet}, {Aerts}, {Goossens},
  {Saesen}, {Cuypers}, {Yakut}, {Scuflaire}, {Dupret}, {Uytterhoeven}, {van
  Winckel}, {Raskin}, {Davignon}, {Le Guillou}, {van Malderen}, {Reyniers},
  {Acke}, {De Meester}, {Vanautgaerden}, {Vandenbussche}, {Verhoelst},
  {Waelkens}, {Deroo}, {Reyniers}, {Ausseloos}, {Broeders},
  {Daszy{\'n}ska-Daszkiewicz}, {Debosscher}, {De Ruyter}, {Lefever}, {Decin},
  {Kolenberg}, {Mazumdar}, {van Kerckhoven}, {De Ridder}, {Drummond}, {Barban},
  {Vanhollebeke}, {Maas}, \& {Decin}}]{DeCat2007}
{De Cat}, P., {Briquet}, M., {Aerts}, C., {et~al.} 2007, \aap, 463, 243 (31)

\bibitem[{{De Cat} {et~al.}(2005){De Cat}, {Briquet},
  {Daszy{\'n}ska-Daszkiewicz}, {Dupret}, {De Ridder}, {Scuflaire}, \&
  {Aerts}}]{DeCat2005}
{De Cat}, P., {Briquet}, M., {Daszy{\'n}ska-Daszkiewicz}, J., {et~al.} 2005,
  \aap, 432, 1013 (8)

\bibitem[{{Degroote} {et~al.}(2012){Degroote}, {Aerts}, {Michel}, {Briquet},
  {P{\'a}pics}, {Amado}, {Mathias}, {Poretti}, {Rainer}, {Lombaert}, {Hillen},
  {Morel}, {Auvergne}, {Baglin}, {Baudin}, {Catala}, \&
  {Samadi}}]{Degroote2012}
{Degroote}, P., {Aerts}, C., {Michel}, E., {et~al.} 2012, \aap, 542, A88 (42)

\bibitem[{{Degroote} {et~al.}(2010){Degroote}, {Briquet}, {Auvergne},
  {Sim{\'o}n-D{\'{\i}}az}, {Aerts}, {Noels}, {Rainer}, {Hareter}, {Poretti},
  {Mahy}, {Oreiro}, {Vu{\v c}kovi{\'c}}, {Smolders}, {Baglin}, {Baudin},
  {Catala}, {Michel}, \& {Samadi}}]{Degroote2010}
{Degroote}, P., {Briquet}, M., {Auvergne}, M., {et~al.} 2010, \aap, 519, A38
  (32)

\bibitem[{{Desmet} {et~al.}(2009){Desmet}, {Briquet}, {Thoul}, {Zima}, {De
  Cat}, {Handler}, {Ilyin}, {Kambe}, {Krzesinski}, {Lehmann}, {Masuda},
  {Mathias}, {Mkrtichian}, {Telting}, {Uytterhoeven}, {Yang}, \&
  {Aerts}}]{Desmet2009}
{Desmet}, M., {Briquet}, M., {Thoul}, A., {et~al.} 2009, \mnras, 396, 1460 (87)

\bibitem[{{Donati} {et~al.}(2006){Donati}, {Howarth}, {Jardine}, {Petit},
  {Catala}, {Landstreet}, {Bouret}, {Alecian}, {Barnes}, {Forveille},
  {Paletou}, \& {Manset}}]{Donati2006}
{Donati}, J.-F., {Howarth}, I.~D., {Jardine}, M.~M., {et~al.} 2006, \mnras,
  370, 629 (66)

\bibitem[{{Fourtune-Ravard} {et~al.}(2011){Fourtune-Ravard}, {Wade},
  {Marcolino}, {Shultz}, {Grunhut}, {Henrichs}, \& {Henrichs}}]{Fourtune2011}
{Fourtune-Ravard}, C., {Wade}, G.~A., {Marcolino}, W.~L.~F., {et~al.} 2011, in
  IAU Symposium, Vol. 272, IAU Symposium, ed. C.~{Neiner}, G.~{Wade},
  G.~{Meynet}, \& G.~{Peters}, 180 (37)

\bibitem[{{Fraser} {et~al.}(2010){Fraser}, {Dufton}, {Hunter}, \&
  {Ryans}}]{Fraser2010}
{Fraser}, M., {Dufton}, P.~L., {Hunter}, I., \& {Ryans}, R.~S.~I. 2010, \mnras,
  404, 1306 (51)

\bibitem[{{Gilks} {et~al.}(1996){Gilks}, {Richardson}, \&
  {Spiegelhalter}}]{Gilks1996}
{Gilks}, W.~R., {Richardson}, S., \& {Spiegelhalter}, D.~J. 1996, {Markov Chain
  Monte Carlo in Practice, Chapman \& Hall, London}

\bibitem[{{Grunhut} {et~al.}(2012){Grunhut}, {Rivinius}, {Wade}, {Townsend},
  {Marcolino}, {Bohlender}, {Szeifert}, {Petit}, {Matthews}, {Rowe}, {Moffat},
  {Kallinger}, {Kuschnig}, {Guenther}, {Rucinski}, {Sasselov}, \&
  {Weiss}}]{Grunhut2012}
{Grunhut}, J.~H., {Rivinius}, T., {Wade}, G.~A., {et~al.} 2012, \mnras, 419,
  1610 (48)

\bibitem[{{Guinan} {et~al.}(2000){Guinan}, {Ribas}, {Fitzpatrick},
  {Gim{\'e}nez}, {Jordi}, {McCook}, \& {Popper}}]{Guinan2000}
{Guinan}, E.~F., {Ribas}, I., {Fitzpatrick}, E.~L., {et~al.} 2000, \apj, 544,
  409

\bibitem[{{Handler} {et~al.}(2006){Handler}, {Jerzykiewicz},
  {Rodr{\'{\i}}guez}, {Uytterhoeven}, {Amado}, {Dorokhova}, {Dorokhov},
  {Poretti}, {Sareyan}, {Parrao}, {Lorenz}, {Zsuffa}, {Drummond},
  {Daszy{\'n}ska-Daszkiewicz}, {Verhoelst}, {De Ridder}, {Acke}, {Bourge},
  {Movchan}, {Garrido}, {Papar{\'o}}, {Sahin}, {Antoci}, {Udovichenko},
  {Csorba}, {Crowe}, {Berkey}, {Stewart}, {Terry}, {Mkrtichian}, \&
  {Aerts}}]{Handler2006}
{Handler}, G., {Jerzykiewicz}, M., {Rodr{\'{\i}}guez}, E., {et~al.} 2006,
  \mnras, 365, 327 (88)

\bibitem[{{Handler} {et~al.}(2009){Handler}, {Matthews}, {Eaton},
  {Daszy{\'n}ska-Daszkiewicz}, {Kuschnig}, {Lehmann}, {Rodr{\'{\i}}guez},
  {Pamyatnykh}, {Zdravkov}, {Lenz}, {Costa}, {D{\'{\i}}az-Fraile}, {Sota},
  {Kwiatkowski}, {Schwarzenberg-Czerny}, {Borczyk}, {Dimitrov}, {Fagas},
  {Kami{\'n}ski}, {Ro{\.z}ek}, {van Wyk}, {Pollard}, {Kilmartin}, {Weiss},
  {Guenther}, {Moffat}, {Rucinski}, {Sasselov}, \& {Walker}}]{Handler2009}
{Handler}, G., {Matthews}, J.~M., {Eaton}, J.~A., {et~al.} 2009, \apjl, 698,
  L56 (3)

\bibitem[{{Handler} {et~al.}(2004){Handler}, {Shobbrook}, {Jerzykiewicz},
  {Krisciunas}, {Tshenye}, {Rodr{\'{\i}}guez}, {Costa}, {Zhou}, {Medupe},
  {Phorah}, {Garrido}, {Amado}, {Papar{\'o}}, {Zsuffa}, {Ramokgali}, {Crowe},
  {Purves}, {Avila}, {Knight}, {Brassfield}, {Kilmartin}, \&
  {Cottrell}}]{Handler2004}
{Handler}, G., {Shobbrook}, R.~R., {Jerzykiewicz}, M., {et~al.} 2004, \mnras,
  347, 454 (12)

\bibitem[{{Handler} {et~al.}(2005){Handler}, {Shobbrook}, \&
  {Mokgwetsi}}]{Handler2005}
{Handler}, G., {Shobbrook}, R.~R., \& {Mokgwetsi}, T. 2005, \mnras, 362, 612
  (69)

\bibitem[{{Handler} {et~al.}(2012){Handler}, {Shobbrook}, {Uytterhoeven},
  {Briquet}, {Neiner}, {Tshenye}, {Ngwato}, {van Winckel}, {Guggenberger},
  {Raskin}, {Rodr{\'{\i}}guez}, {Mazumdar}, {Barban}, {Lorenz},
  {Vandenbussche}, {{\c S}ahin}, {Medupe}, \& {Aerts}}]{Handler2012}
{Handler}, G., {Shobbrook}, R.~R., {Uytterhoeven}, K., {et~al.} 2012, \mnras,
  424, 2380 (72)

\bibitem[{{Henrichs} {et~al.}(2013){Henrichs}, {de Jong}, {Verdugo}, {Schnerr},
  {Neiner}, {Donati}, {Catala}, {Shorlin}, {Wade}, {Veen}, {Nichols}, {Damen},
  {Talavera}, {Hill}, {Kaper}, {Tijani}, {Geers}, {Wiersema}, {Plaggenborg}, \&
  {Rygl}}]{Henrichs2013}
{Henrichs}, H.~F., {de Jong}, J.~A., {Verdugo}, E., {et~al.} 2013, \aap, 555,
  A46 (83)

\bibitem[{{Henrichs} {et~al.}(2012){Henrichs}, {Kolenberg}, {Plaggenborg},
  {Marsden}, {Waite}, {Landstreet}, {Wade}, {Grunhut}, \&
  {Oksala}}]{Henrichs2012}
{Henrichs}, H.~F., {Kolenberg}, K., {Plaggenborg}, B., {et~al.} 2012, \aap,
  545, A119 (60)

\bibitem[{{Henrichs} {et~al.}(2009){Henrichs}, {Neiner}, {Schnerr}, {Verdugo},
  {Alecian}, {Catala}, {Cochard}, {Guti{\'e}rrez}, {Huat}, {Silvester}, \&
  {Thizy}}]{Henrichs2009}
{Henrichs}, H.~F., {Neiner}, C., {Schnerr}, R.~S., {et~al.} 2009, in IAU
  Symposium, Vol. 259, {Cosmic Magnetic Fields: From Planets, to Stars and
  Galaxies}, ed. K.~G. {Strassmeier}, A.~G. {Kosovichev}, \& J.~E. {Beckman},
  393 (84)

\bibitem[{{Hermans} {et~al.}(1990){Hermans}, {Goossens}, \&
  {Kerner}}]{Hermans1990}
{Hermans}, D., {Goossens}, M., \& {Kerner}, W. 1990, \aap, 231, 259

\bibitem[{{Herrero} {et~al.}(1992){Herrero}, {Kudritzki}, {Vilchez}, {Kunze},
  {Butler}, \& {Haser}}]{Herrero1992}
{Herrero}, A., {Kudritzki}, R.~P., {Vilchez}, J.~M., {et~al.} 1992, \aap, 261,
  209

\bibitem[{{Hubrig} {et~al.}(2009){Hubrig}, {Briquet}, {De Cat}, {Sch{\"o}ller},
  {Morel}, \& {Ilyin}}]{Hubrig2009}
{Hubrig}, S., {Briquet}, M., {De Cat}, P., {et~al.} 2009, Astronomische
  Nachrichten, 330, 317 (39)

\bibitem[{{Hubrig} {et~al.}(2006){Hubrig}, {Briquet}, {Sch{\"o}ller}, {De Cat},
  {Mathys}, \& {Aerts}}]{Hubrig2006}
{Hubrig}, S., {Briquet}, M., {Sch{\"o}ller}, M., {et~al.} 2006, \mnras, 369,
  L61 (58)

\bibitem[{{Hunter} {et~al.}(2008){Hunter}, {Brott}, {Lennon}, {Langer},
  {Dufton}, {Trundle}, {Smartt}, {de Koter}, {Evans}, \& {Ryans}}]{Hunter2008}
{Hunter}, I., {Brott}, I., {Lennon}, D.~J., {et~al.} 2008, \apjl, 676, L29

\bibitem[{{Johnson} \& {Wichern}(2000)}]{JohnsonWichern2000}
{Johnson}, R.~A. \& {Wichern}, D.~W. 2000, {Applied Multivariate Statistical
  Analysis, 4th Edition, Englewood Cliffs, Prentice-Hall}

\bibitem[{{Kaper} {et~al.}(1996){Kaper}, {Henrichs}, {Nichols}, {Snoek},
  {Volten}, \& {Zwarthoed}}]{Kaper1996}
{Kaper}, L., {Henrichs}, H.~F., {Nichols}, J.~S., {et~al.} 1996, \aaps, 116,
  257 (23)

\bibitem[{{Kochukhov} {et~al.}(2011){Kochukhov}, {Lundin}, {Romanyuk}, \&
  {Kudryavtsev}}]{Kochukhov2011}
{Kochukhov}, O., {Lundin}, A., {Romanyuk}, I., \& {Kudryavtsev}, D. 2011, \apj,
  726, 24 (25)

\bibitem[{{Koen} \& {Eyer}(2002)}]{Koen2002}
{Koen}, C. \& {Eyer}, L. 2002, \mnras, 331, 45 (61)

\bibitem[{{Krzanowski}(1988)}]{Krzanowski1988}
{Krzanowski}, W.~J. 1988, {Principles of Multivariate Analysis, Clarendon
  Press, Oxford}

\bibitem[{{Langer}(1992)}]{Langer1992}
{Langer}, N. 1992, \aap, 265, L17

\bibitem[{{Lefever} {et~al.}(2010){Lefever}, {Puls}, {Morel}, {Aerts}, {Decin},
  \& {Briquet}}]{Lefever2010}
{Lefever}, K., {Puls}, J., {Morel}, T., {et~al.} 2010, \aap, 515, A74 (43)

\bibitem[{{Leone} {et~al.}(2010){Leone}, {Bohlender}, {Bolton}, {Buemi},
  {Catanzaro}, {Hill}, \& {Stift}}]{Leone2010}
{Leone}, F., {Bohlender}, D.~A., {Bolton}, C.~T., {et~al.} 2010, \mnras, 401,
  2739 (18)

\bibitem[{{Leone} \& {Manfre}(1997)}]{Leone1997}
{Leone}, F. \& {Manfre}, M. 1997, \aap, 320, 257 (73)

\bibitem[{{Little} \& {Rubin}(2002)}]{Little2002}
{Little}, R.~J.~A. \& {Rubin}, D.~B. 2002, {Statistical Analysis with Missing
  Data, John Wiley \& Sons, New York}

\bibitem[{{Lyubimkov} {et~al.}(2002){Lyubimkov}, {Rachkovskaya}, {Rostopchin},
  \& {Lambert}}]{Lyubimkov2002}
{Lyubimkov}, L.~S., {Rachkovskaya}, T.~M., {Rostopchin}, S.~I., \& {Lambert},
  D.~L. 2002, \mnras, 333, 9 (80)

\bibitem[{{Maeder}(2009)}]{Maeder2009}
{Maeder}, A. 2009, {Physics, Formation and Evolution of Rotating Stars,
  Astronomy and Astrophysics Library, Springer Berlin Heidelberg}

\bibitem[{{Martins} {et~al.}(2012{\natexlab{a}}){Martins}, {Escolano}, {Wade},
  {Donati}, {Bouret}, \& {Mimes Collaboration}}]{Martins2012b}
{Martins}, F., {Escolano}, C., {Wade}, G.~A., {et~al.} 2012{\natexlab{a}},
  \aap, 538, A29 (34)

\bibitem[{{Martins} {et~al.}(2012{\natexlab{b}}){Martins}, {Mahy}, {Hillier},
  \& {Rauw}}]{Martins2012a}
{Martins}, F., {Mahy}, L., {Hillier}, D.~J., \& {Rauw}, G. 2012{\natexlab{b}},
  \aap, 538, A39 (1)

\bibitem[{{Mathis}(2013)}]{Mathis2013bis}
{Mathis}, S. 2013, {Habilitation Thesis, Universit\'e Paris XI Orsay, France}

\bibitem[{{Mathis} {et~al.}(2013){Mathis}, {Decressin}, {Eggenberger}, \&
  {Charbonnel}}]{Mathis2013}
{Mathis}, S., {Decressin}, T., {Eggenberger}, P., \& {Charbonnel}, C. 2013,
  \aap, 558, A11

\bibitem[{{Mazumdar} {et~al.}(2006){Mazumdar}, {Briquet}, {Desmet}, \&
  {Aerts}}]{Mazumdar2006}
{Mazumdar}, A., {Briquet}, M., {Desmet}, M., \& {Aerts}, C. 2006, \aap, 459,
  589 (29)

\bibitem[{{Meynet} {et~al.}(2011){Meynet}, {Eggenberger}, \&
  {Maeder}}]{Meynet2011}
{Meynet}, G., {Eggenberger}, P., \& {Maeder}, A. 2011, \aap, 525, L11

\bibitem[{{Mikul{\'a}{\v s}ek} {et~al.}(2008){Mikul{\'a}{\v s}ek}, {Krti{\v
  c}ka}, {Henry}, {Zverko}, {{\v Z}i{\v z}{\aa}ovsk{\'y}}, {Bohlender},
  {Romanyuk}, {Jan{\'{\i}}k}, {Bo{\v z}i{\'c}}, {Kor{\v c}{\'a}kov{\'a}},
  {Zejda}, {Iliev}, {{\v S}koda}, {{\v S}lechta}, {Gr{\'a}f}, {Netolick{\'y}},
  \& {Ceniga}}]{Mikulasek2008}
{Mikul{\'a}{\v s}ek}, Z., {Krti{\v c}ka}, J., {Henry}, G.~W., {et~al.} 2008,
  \aap, 485, 585 (24)

\bibitem[{{Molenberghs} \& {Kenward}(2007)}]{Molenberghs2007}
{Molenberghs}, G. \& {Kenward}, M.~G. 2007, {Missing Data in Clinical Studies,
  John Wiley \& Sons, Chichester}

\bibitem[{{Morel} {et~al.}(2008){Morel}, {Hubrig}, \& {Briquet}}]{Morel2008}
{Morel}, T., {Hubrig}, S., \& {Briquet}, M. 2008, \aap, 481, 453 (28)

\bibitem[{{Neiner} {et~al.}(2012{\natexlab{a}}){Neiner}, {Alecian}, {Briquet},
  {Floquet}, {Fr{\'e}mat}, {Martayan}, {Thizy}, \& {Mimes
  Collaboration}}]{Neiner2012b}
{Neiner}, C., {Alecian}, E., {Briquet}, M., {et~al.} 2012{\natexlab{a}}, \aap,
  537, A148 (71)

\bibitem[{{Neiner} {et~al.}(2012{\natexlab{b}}){Neiner}, {Floquet}, {Samadi},
  {Espinosa Lara}, {Fr{\'e}mat}, {Mathis}, {Leroy}, {de Batz}, {Rainer},
  {Poretti}, {Mathias}, {Guarro Fl{\'o}}, {Buil}, {Ribeiro}, {Alecian},
  {Andrade}, {Briquet}, {Diago}, {Emilio}, {Fabregat}, {Guti{\'e}rrez-Soto},
  {Hubert}, {Janot-Pacheco}, {Martayan}, {Semaan}, {Suso}, \&
  {Zorec}}]{Neiner2012}
{Neiner}, C., {Floquet}, M., {Samadi}, R., {et~al.} 2012{\natexlab{b}}, \aap,
  546, A47

\bibitem[{{Neiner} {et~al.}(2003){Neiner}, {Geers}, {Henrichs}, {Floquet},
  {Fr{\'e}mat}, {Hubert}, {Preuss}, \& {Wiersema}}]{Neiner2003}
{Neiner}, C., {Geers}, V.~C., {Henrichs}, H.~F., {et~al.} 2003, \aap, 406, 1019
  (6)

\bibitem[{{Neiner} {et~al.}(2012{\natexlab{c}}){Neiner}, {Landstreet},
  {Alecian}, {Owocki}, {Kochukhov}, {Bohlender}, \& {MiMeS
  Collaboration}}]{Neiner2012a}
{Neiner}, C., {Landstreet}, J.~D., {Alecian}, E., {et~al.} 2012{\natexlab{c}},
  \aap, 546, A44 (53)

\bibitem[{{Neter} {et~al.}(1996){Neter}, {Nachtsheim}, {Wasserman}, \&
  {Kutner}}]{Neter1996}
{Neter}, J., {Nachtsheim}, C.~J., {Wasserman}, W., \& {Kutner}, M.~H. 1996,
  {Applied Linear Statistical Models, Richard D.\ Irwin Inc., Homewood,
  Illinois}

\bibitem[{{Nieva} \& {Przybilla}(2012)}]{Nieva2012}
{Nieva}, M.-F. \& {Przybilla}, N. 2012, \aap, 539, A143 (5)

\bibitem[{{Oksala} {et~al.}(2012){Oksala}, {Wade}, {Townsend}, {Owocki},
  {Kochukhov}, {Neiner}, {Alecian}, \& {Grunhut}}]{Oksala2012}
{Oksala}, M.~E., {Wade}, G.~A., {Townsend}, R.~H.~D., {et~al.} 2012, \mnras,
  419, 959 (22)

\bibitem[{{Pamyatnykh} {et~al.}(2004){Pamyatnykh}, {Handler}, \&
  {Dziembowski}}]{Pamyatnykh2004}
{Pamyatnykh}, A.~A., {Handler}, G., \& {Dziembowski}, W.~A. 2004, \mnras, 350,
  1022 (11)

\bibitem[{{P{\'a}pics} {et~al.}(2011){P{\'a}pics}, {Briquet}, {Auvergne},
  {Aerts}, {Degroote}, {Niemczura}, {Vu{\v c}kovi{\'c}}, {Smolders}, {Poretti},
  {Rainer}, {Hareter}, {Baglin}, {Baudin}, {Catala}, {Michel}, \&
  {Samadi}}]{Papics2011}
{P{\'a}pics}, P.~I., {Briquet}, M., {Auvergne}, M., {et~al.} 2011, \aap, 528,
  A123 (45)

\bibitem[{{P{\'a}pics} {et~al.}(2012){P{\'a}pics}, {Briquet}, {Baglin},
  {Poretti}, {Aerts}, {Degroote}, {Tkachenko}, {Morel}, {Zima}, {Niemczura},
  {Rainer}, {Hareter}, {Baudin}, {Catala}, {Michel}, {Samadi}, \&
  {Auvergne}}]{Papics2012}
{P{\'a}pics}, P.~I., {Briquet}, M., {Baglin}, A., {et~al.} 2012, \aap, 542, A55
  (26)

\bibitem[{{Paxton} {et~al.}(2011){Paxton}, {Bildsten}, {Dotter}, {Herwig},
  {Lesaffre}, \& {Timmes}}]{Paxton2011}
{Paxton}, B., {Bildsten}, L., {Dotter}, A., {et~al.} 2011, \apjs, 192, 3

\bibitem[{{Paxton} {et~al.}(2013){Paxton}, {Cantiello}, {Arras}, {Bildsten},
  {Brown}, {Dotter}, {Mankovich}, {Montgomery}, {Stello}, {Timmes}, \&
  {Townsend}}]{Paxton2013}
{Paxton}, B., {Cantiello}, M., {Arras}, P., {et~al.} 2013, \apjs, 208, 4

\bibitem[{{Perryman}(1997)}]{Perryman1997}
{Perryman}, M.~A.~C., ed. 1997, ESA Special Publication, Vol. 1200, {The
  HIPPARCOS and TYCHO catalogues. Astrometric and photometric star catalogues
  derived from the ESA HIPPARCOS Space Astrometry Mission}, 1 (2)

\bibitem[{{Petit} {et~al.}(2011){Petit}, {Massa}, {Marcolino}, {Wade},
  {Ignace}, \& {Mimes Collaboration}}]{Petit2011}
{Petit}, V., {Massa}, D.~L., {Marcolino}, W.~L.~F., {et~al.} 2011, \mnras, 412,
  L45 (52)

\bibitem[{{Petit} {et~al.}(2013){Petit}, {Owocki}, {Wade}, {Cohen},
  {Sundqvist}, {Gagn{\'e}}, {Ma{\'{\i}}z Apell{\'a}niz}, {Oksala}, {Bohlender},
  {Rivinius}, {Henrichs}, {Alecian}, {Townsend}, {ud-Doula}, \& {MiMeS
  Collaboration}}]{Petit2013}
{Petit}, V., {Owocki}, S.~P., {Wade}, G.~A., {et~al.} 2013, \mnras, 429, 398
  (14)

\bibitem[{{Petit} {et~al.}(2008){Petit}, {Wade}, {Drissen}, {Montmerle}, \&
  {Alecian}}]{Petit2008}
{Petit}, V., {Wade}, G.~A., {Drissen}, L., {Montmerle}, T., \& {Alecian}, E.
  2008, \mnras, 387, L23 (20)

\bibitem[{{Potter} {et~al.}(2012){Potter}, {Chitre}, \& {Tout}}]{Potter2012}
{Potter}, A.~T., {Chitre}, S.~M., \& {Tout}, C.~A. 2012, \mnras, 424, 2358

\bibitem[{{Przybilla} {et~al.}(2010){Przybilla}, {Firnstein}, {Nieva},
  {Meynet}, \& {Maeder}}]{Przybilla2010}
{Przybilla}, N., {Firnstein}, M., {Nieva}, M.~F., {Meynet}, G., \& {Maeder}, A.
  2010, \aap, 517, A38

\bibitem[{{Rauer} {et~al.}(2013){Rauer}, {Catala}, {Aerts}, {Appourchaux},
  {Benz}, {Brandeker}, {Christensen-Dalsgaard}, {Deleuil}, {Gizon},
  {G{\"u}del}, {Janot-Pacheco}, {Mas-Hesse}, {Pagano}, {Piotto}, {Pollacco},
  {Santos}, {Smith}, {-C.}, {Su{\'a}rez}, {Szab{\'o}}, {Udry}, {Adibekyan},
  {Alibert}, {Almenara}, {Amaro-Seoane}, {Ammler-von Eiff}, {Antonello},
  {Ball}, {Barnes}, {Baudin}, {Belkacem}, {Bergemann}, {Birch}, {Boisse},
  {Bonomo}, {Borsa}, {Brand{\~a}o}, {Brocato}, {Brun}, {Burleigh}, {Burston},
  {Cabrera}, {Cassisi}, {Chaplin}, {Charpinet}, {Chiappini}, {Csizmadia},
  {Cunha}, {Damasso}, {Davies}, {Deeg}, {de Oliveira Fialho}, {D{\'{\i}}az},
  {Dreizler}, {Dreyer}, {Eggenberger}, {Ehrenreich}, {Eigm{\"u}ller},
  {Erikson}, {Farmer}, {Feltzing}, {Figueira}, {Forveille}, {Fridlund},
  {Garc{\'{\i}}a}, {Giuffrida}, {Godolt}, {Gomes da Silva}, {Goupil},
  {Granzer}, {Grenfell}, {Grotsch-Noels}, {G{\"u}nther}, {Haswell}, {Hatzes},
  {H{\'e}brard}, {Hekker}, {Helled}, {Heng}, {Jenkins}, {Khodachenko},
  {Kislyakova}, {Kley}, {Kolb}, {Krivova}, {Kupka}, {Lammer}, {Lanza},
  {Lebreton}, {Magrin}, {Marcos-Arenal}, {Marrese}, {Marques}, {Martins},
  {Mathis}, {Mathur}, {Messina}, {Miglio}, {Montalban}, {Montalto}, {Monteiro},
  {Moradi}, {Moravveji}, {Mordasini}, {Morel}, {Mortier}, {Nascimbeni},
  {Nielsen}, {Noack}, {Norton}, {Ofir}, {Oshagh}, {Ouazzani}, {P{\'a}pics},
  {Parro}, {Petit}, {Plez}, {Poretti}, {Quirrenbach}, {Ragazzoni}, {Raimondo},
  {Rainer}, {Reese}, {Redmer}, {Reffert}, {Rojas-Ayala}, {Roxburgh}, {Solanki},
  {Salmon}, {Santerne}, {Schneider}, {Schou}, {Schuh}, {Schunker},
  {Silva-Valio}, {Silvotti}, {Skillen}, {Snellen}, {Sohl}, {Sousa}, {Sozzetti},
  {Stello}, {Strassmeier}, {{\v S}vanda}, {Szab{\'o}}, {Tkachenko}, {Valencia},
  {van Grootel}, {Vauclair}, {Ventura}, {Wagner}, {Walton}, {Weingrill},
  {Werner}, {Wheatley}, \& {Zwintz}}]{Rauer2013}
{Rauer}, H., {Catala}, C., {Aerts}, C., {et~al.} 2013, Experimental Astronomy,
  submitted (arXiv1310.0696)

\bibitem[{{Rivero Gonz{\'a}lez} {et~al.}(2012){Rivero Gonz{\'a}lez}, {Puls},
  {Najarro}, \& {Brott}}]{Rivero2012}
{Rivero Gonz{\'a}lez}, J.~G., {Puls}, J., {Najarro}, F., \& {Brott}, I. 2012,
  \aap, 537, A79

\bibitem[{{Rivinius} {et~al.}(2003){Rivinius}, {Stahl}, {Baade}, \&
  {Kaufer}}]{Rivinius2003}
{Rivinius}, T., {Stahl}, O., {Baade}, D., \& {Kaufer}, A. 2003, Information
  Bulletin on Variable Stars, 5397, 1 (49)

\bibitem[{{Rivinius} {et~al.}(2013){Rivinius}, {Townsend}, {Kochukhov}, {{\v
  S}tefl}, {Baade}, {Barrera}, \& {Szeifert}}]{Rivinius2013}
{Rivinius}, T., {Townsend}, R.~H.~D., {Kochukhov}, O., {et~al.} 2013, \mnras,
  429, 177 (77)

\bibitem[{{Robert} \& {Casella}(2004)}]{RobertCasella2004}
{Robert}, C.~P. \& {Casella}, G. 2004, {Monte Carlo Statistical Methods,
  Springer, New York}

\bibitem[{{Rogers} {et~al.}(2013){Rogers}, {Lin}, {McElwaine}, \&
  {Lau}}]{Rogers2013}
{Rogers}, T.~M., {Lin}, D.~N.~C., {McElwaine}, J.~N., \& {Lau}, H.~H.~B. 2013,
  \apj, 772, 21

\bibitem[{{Romanyuk} \& {Kudryavtsev}(2008)}]{Romanyuk2008}
{Romanyuk}, I.~I. \& {Kudryavtsev}, D.~O. 2008, Astrophysical Bulletin, 63, 139
  (17)

\bibitem[{{Rubin}(1976)}]{Rubin1976}
{Rubin}, D.~B. 1976, Biometrika, 63, 581

\bibitem[{{Rubin}(1987)}]{Rubin1987}
{Rubin}, D.~B. 1987, {Multiple Imputation for Nonresponse in Surveys, John
  Wiley \& Sons, New York}

\bibitem[{{Saesen} {et~al.}(2006){Saesen}, {Briquet}, \& {Aerts}}]{Saesen2006}
{Saesen}, S., {Briquet}, M., \& {Aerts}, C. 2006, Communications in
  Asteroseismology, 147, 109

\bibitem[{{Schafer}(1997)}]{Schafer1997}
{Schafer}, J.~L. 1997, {Analysis of Incomplete Multivariate Data, Chapman \&
  Hall, London}

\bibitem[{{Shobbrook} {et~al.}(2006){Shobbrook}, {Handler}, {Lorenz}, \&
  {Mogorosi}}]{Shobbrook2006}
{Shobbrook}, R.~R., {Handler}, G., {Lorenz}, D., \& {Mogorosi}, D. 2006,
  \mnras, 369, 171 (44)

\bibitem[{{Shultz} {et~al.}(2012){Shultz}, {Wade}, {Grunhut}, {Bagnulo},
  {Landstreet}, {Neiner}, {Alecian}, {Hanes}, \& {MiMeS
  Collaboration}}]{Shultz2012}
{Shultz}, M., {Wade}, G.~A., {Grunhut}, J., {et~al.} 2012, \apj, 750, 2 (9)

\bibitem[{{Silvester} {et~al.}(2009){Silvester}, {Neiner}, {Henrichs}, {Wade},
  {Petit}, {Alecian}, {Huat}, {Martayan}, {Power}, \& {Thizy}}]{Silvester2009}
{Silvester}, J., {Neiner}, C., {Henrichs}, H.~F., {et~al.} 2009, \mnras, 398,
  1505 (4)

\bibitem[{{Silvester} {et~al.}(2012){Silvester}, {Wade}, {Kochukhov},
  {Bagnulo}, {Folsom}, \& {Hanes}}]{Silvester2012}
{Silvester}, J., {Wade}, G.~A., {Kochukhov}, O., {et~al.} 2012, \mnras, 426,
  1003 (13)

\bibitem[{{Sim{\'o}n-D{\'{\i}}az} {et~al.}(2010){Sim{\'o}n-D{\'{\i}}az},
  {Herrero}, {Uytterhoeven}, {Castro}, {Aerts}, \& {Puls}}]{SimonDiaz2010}
{Sim{\'o}n-D{\'{\i}}az}, S., {Herrero}, A., {Uytterhoeven}, K., {et~al.} 2010,
  \apjl, 720, L174 (85)

\bibitem[{{Stankov} \& {Handler}(2005)}]{Stankov2005}
{Stankov}, A. \& {Handler}, G. 2005, \apjs, 158, 193 (30)

\bibitem[{{Talon} {et~al.}(1997){Talon}, {Zahn}, {Maeder}, \&
  {Meynet}}]{Talon1997}
{Talon}, S., {Zahn}, J.-P., {Maeder}, A., \& {Meynet}, G. 1997, \aap, 322, 209

\bibitem[{{Telting} {et~al.}(1997){Telting}, {Aerts}, \&
  {Mathias}}]{Telting1997}
{Telting}, J.~H., {Aerts}, C., \& {Mathias}, P. 1997, \aap, 322, 493 (82)

\bibitem[{{Thoul} {et~al.}(2003){Thoul}, {Aerts}, {Dupret}, {Scuflaire},
  {Korotin}, {Egorova}, {Andrievsky}, {Lehmann}, {Briquet}, {De Ridder}, \&
  {Noels}}]{Thoul2003}
{Thoul}, A., {Aerts}, C., {Dupret}, M.~A., {et~al.} 2003, \aap, 406, 287 (90)

\bibitem[{{Thoul} {et~al.}(2013){Thoul}, {Degroote}, {Catala}, {Aerts},
  {Morel}, {Briquet}, {Hillen}, {Raskin}, {Van Winckel}, {Auvergne}, {Baglin},
  {Baudin}, \& {Michel}}]{Thoul2013}
{Thoul}, A., {Degroote}, P., {Catala}, C., {et~al.} 2013, \aap, 551, A12 (41)

\bibitem[{{Tkachenko} {et~al.}(2013){Tkachenko}, Degroote, {Aerts},
  {Pavlovski}, {Pavlovski}, \& {Pavlovski}}]{Tkachenko2013}
{Tkachenko}, A., Degroote, P., {Aerts}, C., {et~al.} 2013, \mnras, in press
  (arXiv1312.3601) (79)

\bibitem[{{Townsend} {et~al.}(2010){Townsend}, {Oksala}, {Cohen}, {Owocki}, \&
  {ud-Doula}}]{Townsend2010}
{Townsend}, R.~H.~D., {Oksala}, M.~E., {Cohen}, D.~H., {Owocki}, S.~P., \&
  {ud-Doula}, A. 2010, \apjl, 714, L318 (21)

\bibitem[{{Trundle} {et~al.}(2007){Trundle}, {Dufton}, {Hunter}, {Evans},
  {Lennon}, {Smartt}, \& {Ryans}}]{Trundle2007}
{Trundle}, C., {Dufton}, P.~L., {Hunter}, I., {et~al.} 2007, \aap, 471, 625

\bibitem[{{Venn} \& {Lambert}(2005)}]{Venn2005}
{Venn}, K.~A. \& {Lambert}, D.~L. 2005, in Astronomical Society of the Pacific
  Conference Series, Vol. 336, Cosmic Abundances as Records of Stellar
  Evolution and Nucleosynthesis, ed. T.~G. {Barnes}, III \& F.~N. {Bash}, 93

\bibitem[{{Villamariz} {et~al.}(2002){Villamariz}, {Herrero}, {Becker}, \&
  {Butler}}]{Villamariz2002}
{Villamariz}, M.~R., {Herrero}, A., {Becker}, S.~R., \& {Butler}, K. 2002,
  \aap, 388, 940 (86)

\bibitem[{{von Neumann}(1951)}]{vonNeumann1951}
{von Neumann}, J. 1951, Nat.\ Bureau Standards, 12, 36–38

\bibitem[{{Wade} {et~al.}(1997){Wade}, {Bohlender}, {Brown}, {Elkin},
  {Landstreet}, \& {Romanyuk}}]{Wade1997}
{Wade}, G.~A., {Bohlender}, D.~A., {Brown}, D.~N., {et~al.} 1997, \aap, 320,
  172 (78)

\bibitem[{{Wade} {et~al.}(2012){Wade}, {Grunhut}, {Gr{\"a}fener}, {Howarth},
  {Martins}, {Petit}, {Vink}, {Bagnulo}, {Folsom}, {Naz{\'e}}, {Walborn},
  {Townsend}, \& {Evans}}]{Wade2012}
{Wade}, G.~A., {Grunhut}, J., {Gr{\"a}fener}, G., {et~al.} 2012, \mnras, 419,
  2459 (65)

\bibitem[{{Wade} {et~al.}(2011){Wade}, {Howarth}, {Townsend}, {Grunhut},
  {Shultz}, {Bouret}, {Fullerton}, {Marcolino}, {Martins}, {Naz{\'e}}, {Ud
  Doula}, {Walborn}, \& {Donati}}]{Wade2011}
{Wade}, G.~A., {Howarth}, I.~D., {Townsend}, R.~H.~D., {et~al.} 2011, \mnras,
  416, 3160 (81)

\bibitem[{{Yakunin} {et~al.}(2011){Yakunin}, {Romanyuk}, {Kudryavtsev}, \&
  {Semenko}}]{Yakunin2011}
{Yakunin}, I., {Romanyuk}, I., {Kudryavtsev}, D., \& {Semenko}, E. 2011,
  Astronomische Nachrichten, 332, 974 (16)

\end{thebibliography}

\appendix

\begin{table}[t!]
\caption{\label{data}Data set used in this study composed as per 1 September
  2013.} \tablecomments{The object number corresponds to the Henry-Draper (HD)
  number.  The spectral type of the star (SpT) was retrieved from the
    Simbad data base and was used to order the stars.  The rightmost column is a
    flag to indicate spectroscopic binarity (1) or not (0). These three entries
    were not used in the statistical modeling.  The meaning of $X_i$ is given
  in Table~1. The four stars without acceptable imputed values for their missing
  data are indicated in {\it italic}. The data source for each entry is listed
  in brackets beneath the value as an index which corresponds to the same index
  in the reference list except for (33), which 
stands for an unpublished null
  detection by the MiMeS collaboration.}
\tabcolsep=4pt
\begin{tabular}{rrrrrrrrrrrrr}
\hline
\multicolumn{1}{c}{object}&
\multicolumn{1}{r}{SpT}&
\multicolumn{1}{c}{$X_1$}&
\multicolumn{1}{c}{$X_2$}&
\multicolumn{1}{c}{$X_3$}&
\multicolumn{1}{c}{$X_4$}&
\multicolumn{1}{c}{$X_5$}&
\multicolumn{1}{c}{$X_6$}&
\multicolumn{1}{c}{$X_7$}&
\multicolumn{1}{c}{$X_8$}&
\multicolumn{1}{c}{$X_9$}&
\multicolumn{1}{c}{$X_{10}$}&
\multicolumn{1}{c}{SB}\\
\hline
   108& O4-8f?p& 0& 0.000& ---  & $<$4.80& ---  & $<$4.80& 3.10  & 4.544& 3.50&
 8.77&0\\
&& (1) & (1) & --- & (2) & --- & (2) & (1) & (1) & (1) & (1)  \\
 46223&O5e&100& 0.249& 0.755&    0.46& 7.554& 0.17   & 0.00  & 4.633& 4.01& 8.85&0\\
&& (34) & (35) & (35) & (35) & (35) & (35) & (33) & (34) & (34) & (34) & \\
 46150&O5Vf&100& 0.144& 0.055&    0.32& ---  & $<$0.01& 0.00  & 4.623& 4.01& 8.48&0\\
&& (34) & (35) & (35) & (35) & --- & (35) & (33) & (34) & (34) & (34) & \\
148937&O5.5-6f?p& 45& 0.142& ---  &$<$10.00& ---  &$<$10.00& 3.01  & 4.602& 4.00& 8.48&1\\
&& (65) & (65) & --- & (2) & --- & (2) & (65) & (1) & (1) & (1) & \\
 37022&O7V& 24& 0.065& ---  &$<$22.80& ---  &$<$22.80& 3.04  & 4.591& 4.10& 7.82&1\\
&& (1) & (20) & --- & (2) & --- & (2) & (1) & (1) & (1) & (1) & \\
191612&O8fpe&  1& 0.002& ---  & $<$8.00& ---  & $<$8.00& 3.39  & 4.556& 3.75& 8.43&1\\
&& (1) & (81) & --- & (2) & --- & (2) & (81) & (1) & (1) & (1) & \\
 46966&O8V& 50& 0.084& 0.039&    0.89& 8.226&   0.06& 0.00  & 4.544& 3.75& 8.08&0\\
&& (34) & (35) & (35) & (35) & (35) & (35) & (33) & (34) & (34) & (34) & \\
 46149&O8.5V& 30& 0.084& 0.175&    0.20& 3.844&    0.05& 0.00  & 4.556& 3.70& 7.90&1\\
\end{tabular}\end{table}
\setcounter{table}{2}
\begin{table}\caption{Continued.}
\begin{tabular}{rrrrrrrrrrrrr}
\hline
\multicolumn{1}{c}{object}&
\multicolumn{1}{r}{SpT}&
\multicolumn{1}{c}{$X_1$}&
\multicolumn{1}{c}{$X_2$}&
\multicolumn{1}{c}{$X_3$}&
\multicolumn{1}{c}{$X_4$}&
\multicolumn{1}{c}{$X_5$}&
\multicolumn{1}{c}{$X_6$}&
\multicolumn{1}{c}{$X_7$}&
\multicolumn{1}{c}{$X_8$}&
\multicolumn{1}{c}{$X_9$}&
\multicolumn{1}{c}{$X_{10}$}&
\multicolumn{1}{c}{SB}\\
\hline
 57682&O9IV& 15& 0.016& ---  & $<$3.20& ---  & $<$3.20& 2.94  & 4.538& 4.00& 8.11&0\\
&& (1) & (14) & --- & (2) & --- & (2) & (48) & (48) & (48) & (1) & \\
214680&O9V& 16& 0.147&   --- & $<$2.40& 3.258& 5.50& 0.00  & 4.532& 4.10& 7.81&0\\
&& (85) & (23) & --- & (2) & (31) & (31) & (33) & (31) & (31) & (86) & \\
 46202&O9V& 25&  --- & 0.510&    0.11& 4.856&    0.09& 0.00  & 4.525& 4.10& 8.00&0\\
&& (36) & --- & (36) & (36) & (36) & (36)& (33) & (34) & (34) & (34) & \\
 37742&O9Iab:&100& 0.167& ---  & $<$4.40& ---  & $<$4.40& 1.88  & 4.470& 3.25& 7.52&1\\
&& (1) & (23) & --- & (2) & --- & (2) & (1) & (1) & (1) & (1) & \\
149438&B0.2V&  6& 0.024& ---  & $<$1.60& ---  & $<$1.60& 2.70  & 4.491& 4.10& 8.15&0\\
&& (66) & (66) & --- & (2) & --- & (2) & (66) & (1) & (1) & (1) & \\
 51756&B0.5IV& 28& 0.527& ---  & $<$0.01& ---  & $<$0.01& 0.00  & 4.477& 3.75& --- &0\\
&& (45) & (45) & --- & (2) & --- & (2) & (33) & (45) & (45) & --- & \\
111123&B0.5IV& 18& ---  & ---  & $<$4.00& 5.231&    3.10& 0.00  & 4.439& 3.65& 7.61&1\\
&& (55) & --- & --- & (2) & (56) & (56) & (33) & (28) & (28) & (28) & \\ 
 46328&B0.7IV&  9& 0.235& ---  & $<$8.40& 4.772&   16.20&$>3.08$& 4.439& 3.75& 8.00&0\\
&& (28) & (37) & --- & (2) & (38) & (39) & (39) & (28) & (28) & (28) & \\
 44743&B1III& 23& 0.054& ---  & $<$5.60& 3.979&   21.00& 0.00  & 4.380& 3.50& 7.59&0\\
&& (28) & (29) & --- & (2) & (30) & (30) & (4) & (28) & (28) & (28) & \\
 50707&B1Ib& 34& 0.107& ---  & $<$2.40& 5.419&    4.6& 0.00  & 4.415& 3.60& 8.03&0\\
&& (43) & (44) & --- & (2) & (44) & (44) & (9) & (28) & (28) & (28) & \\
 66665&B1V& 10& 0.048& ---  & $<$6.00& ---  & $<$6.00& 2.83  & 4.447& 3.90& --- &0\\
&& (52) & (52) & --- & (2) & --- & (2) & (52) & (52) & (52) & --- & \\
187879&B1III+& 97& 0.129& ---  & $<$0.01& ---  & $<$0.01& ---   & 4.336& 3.10& 7.55&1\\
&& (79) & (79) & --- & (79) & --- & (79) & --- & (79) & (79) & (79) & \\
\end{tabular}\end{table}
\setcounter{table}{2}
\begin{table}\caption{Continued.}
\begin{tabular}{rrrrrrrrrrrrr}
\hline
\multicolumn{1}{c}{object}&
\multicolumn{1}{r}{SpT}&
\multicolumn{1}{c}{$X_1$}&
\multicolumn{1}{c}{$X_2$}&
\multicolumn{1}{c}{$X_3$}&
\multicolumn{1}{c}{$X_4$}&
\multicolumn{1}{c}{$X_5$}&
\multicolumn{1}{c}{$X_6$}&
\multicolumn{1}{c}{$X_7$}&
\multicolumn{1}{c}{$X_8$}&
\multicolumn{1}{c}{$X_9$}&
\multicolumn{1}{c}{$X_{10}$}&
\multicolumn{1}{c}{SB}\\
\hline
 37017&B1.5V& 90& 1.110& ---  & $<$5.60& ---  & $<$5.60&$>$3.78& 4.322& 4.10& --- &1\\
&& (19) & (19) & --- & (2) & --- & (2) & (14) & (19) & (19) & --- & \\
 64740&B1.5Vp&160& 0.752& ---  & $<$2.00& ---  & $<$2.00& 4.20  & 4.380& 4.00&7.89 &0\\
&& (50) & (50) & --- & (2) & --- & (2) & (14) & (14) & (14) & (51) & \\
 74575&B1.5III& 11& ---  & ---  & $<$2.00& ---  & $<$2.00& 0.00  & 4.360& 3.60& 7.92&0\\
&& (5) & --- & --- & (2) & --- & (2) & (9) & (5) & (5) & (5) & \\
180642&B1.5III& 25& 0.075& 0.308&    1.60& 5.487&   38.10& 0.00  & 4.389& 3.45& 8.00&0\\
&& (75) & (75) & (76) & (76) & (76) & (76) & (9) & (76) & (76) & (76) & \\
   886&B2IV&  3& 0.005& 0.682&    1.99& 6.590&    6.59& 0.00  & 4.342& 3.95& 7.76&0\\
&& (3) & (3) & (3) & (3) & (3) & (3) & (4) & (5) & (5) & (5) & \\
{\it 3360}& B2IV&17& 0.186& 0.640&    1.30& ---  & $<$1.00& 2.53  & 4.317& 3.80&
8.23&0\\
&& (6) & (6) & (6) & (2) & --- & (2) & (6) & (5) & (5) & (5) \\
 16582&B2IV&  1& 0.075& 0.318&    0.43& 6.206&   11.62& 0.00  & 4.327& 3.80& 8.23&0\\
&& (7) & (7) & (7) & (7) & (7) & (7) & (8) & (5) & (5) & (5) \\
 29248&B2III&  6& 0.017& 0.432&    3.20& 5.763&   36.90& 0.00  & 4.342& 3.85& 7.93&0\\
&& (11) & (11) & (12) & (12) & (12) & (12) & (4) & (5) & (5) & (5) & \\
 36485&B2V& 32& 0.677& ---  &     ---& ---  & ---    & 4.00  & 4.301& 4.00& --- &1\\
&& (18) & (18) & --- & --- & --- & --- & (18) & (18) & (18) & --- & \\
 37479&B2Vp&170& 0.840& ---  &    --- & ---  &    --- & 3.98  & 4.362& 4.00& --- &0\\
&& (14) & (21) & --- & --- & --- & --- & (22) & (14) & (14) & --- & \\
 37776&B2IV& 95& 0.650& ---  & $<$7.20& ---  & $<$7.20& 4.18  & 4.342& 4.00& --- &0\\
&& (14) & (24) & --- & (2) & --- & (2) & (25) & (14) & (14) & --- & \\
 55522&B2V& 70& 0.366& ---  & $<$7.20& ---  & $<$7.20&$>$3.42& 4.241& 4.20& --- &0\\
&& (46) & (46) & --- & (2) & --- & (2) & (47) & (47) & (47) & --- & \\
\end{tabular}\end{table}
\setcounter{table}{2}
\begin{table}\caption{Continued.}
\begin{tabular}{rrrrrrrrrrrrr}
\hline
\multicolumn{1}{c}{object}&
\multicolumn{1}{r}{SpT}&
\multicolumn{1}{c}{$X_1$}&
\multicolumn{1}{c}{$X_2$}&
\multicolumn{1}{c}{$X_3$}&
\multicolumn{1}{c}{$X_4$}&
\multicolumn{1}{c}{$X_5$}&
\multicolumn{1}{c}{$X_6$}&
\multicolumn{1}{c}{$X_7$}&
\multicolumn{1}{c}{$X_8$}&
\multicolumn{1}{c}{$X_9$}&
\multicolumn{1}{c}{$X_{10}$}&
\multicolumn{1}{c}{SB}\\
\hline
 61068&B2II& 10& ---  & ---  & $<$3.60& 6.010&   19.50& 0.00  & 4.420& 4.15& 8.00&0\\
&& (39) & --- & --- & (2) & (30) & (30) & (39) & (5) & (5) & (5) & \\
 67621&B2IV& 20& 0.279& ---  & $<$3.20& ---  & $<$3.20&$>$2.95& 4.279& 4.00& --- &0\\
&& (14) & (14) & --- & (2) & --- & (2) & (14) & (14) & (14) & --- & \\
{\it 85953}&B2III& 18& 0.020& 0.266&   11.20& ---  & $<$1.00& 0.00  & 4.322& 3.80& 7.66&0\\
&& (39) & (10) & (10) & (10) & --- & (10) & (33) & (28) & (28) & (28) & \\
 96446&B2IIIp&  3& 0.175& 1.149& $<$4.40&10.763& $<$4.40& 3.89  & 4.334& 4.00& 7.42&0\\
127381&B2III& 68& 0.331& ---  & $<$2.00&10.935&    3.10& 2.70  & 4.362& 4.02& 8.26&0\\
&& (60) & (60) & --- & (2) & (61) & (61) & (60) & (60) & (60) & (60) & \\
142184&B2V&290& 1.967& ---  & $<$4.80& ---  & $<$4.80& 4.00  & 4.230& 4.25& --- &0\\
&& (48) & (48) & --- & (2) & --- & (2) & (48) & (48) & (48) & --- & \\
{\it 157056}&B2IV& 31& 0.107& ---  & $<$1.00& 7.116&    9.40&  ---  & 4.398& 4.10& 7.78&1\\
&& (67) & (68) & --- & (69) & (69) & (69) & (58) & (28) & (28) & (28) & \\
163472&B2V& 63& 0.275& ---  & $<$1.00& 7.148&   13.22& 2.60  & 4.352& 3.95& 7.99&0\\
&& (70) & (71) & --- & (72) & (72) & (72) & (71) & (28) & (28) & (28) & \\
182180&B2Vn&310& 1.918& ---  &$<$10.00& ---  &$<$10.00& 4.06  & 4.248& 4.05& --- &0\\
&& (77) & (77) & --- & (2) & --- & (2) & (77) & (77) & (77) & --- & \\
184927&B2V& 14& 0.105& ---  & $<$4.40& ---  & $<$4.40& 3.59  & 4.342& 3.90& --- &0\\
&& (78) & (78) & --- & (2) & --- & (2) & (14) & (14) & (14) & --- & \\
205021&B2IIIev& 25& 0.083& ---  & $<$1.00& 5.250&   37.00& 2.48  & 4.431& 3.75& 8.11&0\\
&& (82) & (83) & --- & (2) & (30) & (30) & (83) & (5) & (5) & (5) & \\
214993&B2III& 36& 0.120& 0.355&    5.00& 5.179&   38.10& 0.00  & 4.389& 3.65& 7.64&0\\
&& (87) & (87) & (88) & (88) & (88) & (88) & (33) & (28) & (28) & (28) & \\
\end{tabular}\end{table}
\setcounter{table}{2}
\begin{table}\caption{Continued.}
\begin{tabular}{rrrrrrrrrrrrr}
\hline
\multicolumn{1}{c}{object}&
\multicolumn{1}{r}{SpT}&
\multicolumn{1}{c}{$X_1$}&
\multicolumn{1}{c}{$X_2$}&
\multicolumn{1}{c}{$X_3$}&
\multicolumn{1}{c}{$X_4$}&
\multicolumn{1}{c}{$X_5$}&
\multicolumn{1}{c}{$X_6$}&
\multicolumn{1}{c}{$X_7$}&
\multicolumn{1}{c}{$X_8$}&
\multicolumn{1}{c}{$X_9$}&
\multicolumn{1}{c}{$X_{10}$}&
\multicolumn{1}{c}{SB}\\
\hline
216916&B2IV& 20& 0.275& ---  & $<$2.00& 5.911&    2.55& 0.00  & 4.362& 3.95& 7.78&1\\
&& (89) & (90) & --- & (91) & (91) & (91) & (33) & (5) & (5) & (5) & \\
 48977&B2.5V& 29& 0.637& 0.517&    2.23& ---  & $<$0.01& 0.00  & 4.301& 4.20& 7.53&0\\
&& (41) & (41) & (41) & (41) & --- & (41) & (33) & (41) & (41) & (41) & \\
208057&B3Ve&104& 0.694& 0.802&    9.00& ---  & $<$7.60&$>$2.70& 4.279& 3.90& --- &0\\
&& (58) & (84) & (31) & (31) & --- & (2) & (15) & (15) & (15) & ---  & \\
{\it 129929}&B3V&  2& 0.012& ---  & $<$1.00& 6.462&   11.80& 0.00  & 4.389& 3.95& 7.73&0\\
&& (62) & (63) & --- & (62) & (62) & (62) & (33) & (28) & (28) & (28) & \\
 50230&B3V&  7& 0.044& 0.684&    1.84& 4.922&    1.24&  ---  & 4.255& 3.80& --- &1\\
&& (42) & (42) & (42) & (42) & (42) & (42) & --- & (42) & (42) & --- & \\
 74560&B3IV& 13& 0.010& 0.645&   14.30& ---  & $<$1.00& 0.00  & 4.210& 4.15& --- &1\\
&& (39) & (10) & (10) & (10) & --- & (2) & (9) & (39) & (39) & --- & \\
 43317&B3IV&110& 1.115& 1.101&    1.44& 4.331&    0.56& 3.08  & 4.230& 4.00& 7.66&0\\
&& (26) & (26) & (26) & (26) & (26) & (26) & (27) & (26) & (26) & (26) & \\
 35298&B3Vw&260& 0.540& ---  & $<$9.60& ---  & $<$9.60&$>$3.95& 4.204& 3.80& --- &0\\
&& (14) & (15) & --- & (2) & --- & (2) & (16) & (14) & (14) & --- & \\
 24587&B5V& 32& 0.426& 1.157&    7.90& ---  & $<$1.00& 0.00  & 4.142& 4.26& --- &1\\
&& (8) & (8) & (8) & (8) & --- & (2) & (9) & (10) & (10) & --- & \\
189775&B5III& 85& 0.384& ---  &$<$10.40& ---  &$<$10.40&$>$3.65& 4.204& 3.80& --- &0\\
&& (80) & (80) & --- & (2) & --- & (2) & (14) & (80) & (80) & --- & \\
176582&B5IV&105& 0.632& ---  & $<$7.20& ---  & $<$7.20& 3.85  & 4.204& 4.00& --- &0\\
&& (74) & (74) & --- & (2) & --- & (2) & (74) & (74) & (74) & --- & \\
142990&B5V&125& 1.021& ---  & $<$4.80& ---  & $<$4.80&$>$3.88& 4.230& 4.20& --- &0\\
&& (64) & (15) & --- & (2) & --- & (2) & (15) & (64) & (64) & --- & \\
\end{tabular}\end{table}
\setcounter{table}{2}
\begin{table}\caption{Continued.}
\begin{tabular}{rrrrrrrrrrrrr}
\hline
\multicolumn{1}{c}{object}&
\multicolumn{1}{r}{SpT}&
\multicolumn{1}{c}{$X_1$}&
\multicolumn{1}{c}{$X_2$}&
\multicolumn{1}{c}{$X_3$}&
\multicolumn{1}{c}{$X_4$}&
\multicolumn{1}{c}{$X_5$}&
\multicolumn{1}{c}{$X_6$}&
\multicolumn{1}{c}{$X_7$}&
\multicolumn{1}{c}{$X_8$}&
\multicolumn{1}{c}{$X_9$}&
\multicolumn{1}{c}{$X_{10}$}&
\multicolumn{1}{c}{SB}\\
\hline
 61556&B5IVn& 70& 0.524& ---  &    --- &   ---&     ---& 3.60  & 4.176& 4.00& --- &0\\
&& (49) & (49) & --- & --- & --- & --- & (14) & (14) & (14) & --- & \\
 46769&B5II& 72& 0.103& ---  & $<$0.01& ---  & $<$0.01& 0.00  & 4.114& 2.70& 8.08&0\\
&& (40) & (40) & --- & (40) & --- & (40) & (40) & (40) & (40) & (40) & \\
 35502&B5V& 80& 1.176& ---  & $<$5.60& ---  & $<$5.60&$>$3.83& 4.204& 3.80& --- &1\\
&& (14) & (14) & --- & (2) & --- & (2) & (17) & (14) & (14) & --- & \\
105382&B6IIIe& 73& 0.772& ---  & $<$7.20& ---  & $<$7.20& 3.36  & 4.241& 4.18& --- &0\\
&& (46) & (46) & --- & (2) & --- & (2) & (54) & (46) & (46) & --- & \\
125823&B7IIIpv& 15& 0.113& ---  &$<$18.80& ---  &$<$18.80&$>$4.15& 4.279& 4.00& 8.10&0\\
&& (59) & (59) & --- & (2) & --- & (2) & (59) & (59) & (59) & (59) & \\
 46005&B8V&150& 2.505& ---  &$<$26.00& ---  &$<$26.00& 0.00  & 4.325& 4.43& --- &0\\
&& (4) & (31) & --- & (2) & --- & (2) & (4) & (31) & (31) & --- & \\
175362&B8IVs& 35& 0.272& ---  &$<$18.00& ---  &$<$18.00&$>$4.32& 4.176& 3.70& 8.95&0\\
&& (73) & (15) & --- & (2) & --- & (2) & (15) & (73) & (73) & (73) & \\
140873&B8III& 69& 1.226& 1.152&   13.40& ---  & $<$1.00& 0.00  & 4.144& 4.35& --- &1\\
&& (8) & (8) & (10) & (10) & --- & (10) & (33) & (57) & (57) & --- & \\
 32633&B9p& 19& 0.156& ---  & $<$5.20& ---  & $<$5.20&$>$3.93& 4.107& 4.17& --- &0\\
&& (13) & (13) & --- & (2) & --- & (2) & (13) & (13) & (13) & --- & \\
123515&B9IV& 15& 0.038& 0.685&   20.90& ---  & $<$1.00& 0.00  & 4.097& 4.29& --- &1\\
&& (57) & (10) & (10) & (10) & --- & (10) & (58) & (10) & (10) & --- & \\
\end{tabular}\end{table}
\setcounter{table}{2}
\begin{table}\caption{Continued.}
\tablerefs{{\tiny
(1) \citet{Martins2012a};
(2) \citet{Perryman1997};
(3) \citet{Handler2009};
(4) \citet{Silvester2009};
(5) \citet{Nieva2012};
(6) \citet{Neiner2003};
(7) \citet{Aerts2006};
(8) \citet{DeCat2005};
(9) \citet{Shultz2012};
(10) \citet{DeCatAerts2002};
(11) \citet{Pamyatnykh2004};
(12) \citet{Handler2004};
(13) \citet{Silvester2012};
(14) \citet{Petit2013};
(15) \citet{Bychkov2005};
(16) \citet{Yakunin2011};
(17) \citet{Romanyuk2008};
(18) \citet{Leone2010};
(19) \citet{Bolton1998};
(20) \citet{Petit2008};
(21) \citet{Townsend2010};
(22) \citet{Oksala2012};
(23) \citet{Kaper1996};
(24) \citet{Mikulasek2008};
(25) \citet{Kochukhov2011};
(26) \citet{Papics2012};
(27) \citet{Briquet2013};
(28) \citet{Morel2008};
(29) \citet{Mazumdar2006};
(30) \citet{Stankov2005};
(31) \citet{DeCat2007};
(32) \citet{Degroote2010};
(33) MiMeS collaboration, 
unpublished null
detection; 
(34) \citet{Martins2012b};
(35) \citet{Blomme2011};
(36) \citet{Briquet2011};
(37) \citet{Fourtune2011};
(38) \citet{Saesen2006};
(39) \citet{Hubrig2009};
(40) \citet{Aerts2013a};\pagebreak
(41) \citet{Thoul2013};
(42) \citet{Degroote2012};
(43) \citet{Lefever2010};
(44) \citet{Shobbrook2006};
(45) \citet{Papics2011};
(46) \citet{Briquet2004};
(47) \citet{Briquet2007a};
(48) \citet{Grunhut2012};
(49) \citet{Rivinius2003};
(50) \citet{Bohlender1990};
(51) \citet{Fraser2010};
(52) \citet{Petit2011};
(53) \citet{Neiner2012a};
(54) \citet{Alecian2011};
(55) \citet{Aerts1998};
(56) \citet{Cuypers2002};
(57) \citet{Aerts1999};
(58) \citet{Hubrig2006};
(59) \citet{Bohlender2010};
(60) \citet{Henrichs2012};
(61) \citet{Koen2002};
(62) \citet{Aerts2004};
(63) \citet{Aerts2003a};
(64) \citet{Cidale2007};
(65) \citet{Wade2012};
(66) \citet{Donati2006};
(67) \citet{Briquet2005};
(68) \citet{Briquet2007b};
(69) \citet{Handler2005};
(70) \citet{Briquet2012};
(71) \citet{Neiner2012b};
(72) \citet{Handler2012};
(73) \citet{Leone1997};
(74) \citet{Bohlender2011};
(75) \citet{Aerts2011};
(76) \citet{Briquet2009}
(77) \citet{Rivinius2013};
(78) \citet{Wade1997};
(79) \citet{Tkachenko2013};
(80) \citet{Lyubimkov2002};
(81) \citet{Wade2011};
(82) \citet{Telting1997};
(83) \citet{Henrichs2013};
(84) \citet{Henrichs2009};
(85) \citet{SimonDiaz2010};
(86) \citet{Villamariz2002};
(87) \citet{Desmet2009};
(88) \citet{Handler2006};
(89) \citet{Aerts2003b};
(90) \citet{Thoul2003};
(91) \citet{Chapellier1995}
}}
\begin{tabular}{rrrrrrrrrrrrr}
\hline
\multicolumn{1}{c}{object}&
\multicolumn{1}{r}{SpT}&
\multicolumn{1}{c}{$X_1$}&
\multicolumn{1}{c}{$X_2$}&
\multicolumn{1}{c}{$X_3$}&
\multicolumn{1}{c}{$X_4$}&
\multicolumn{1}{c}{$X_5$}&
\multicolumn{1}{c}{$X_6$}&
\multicolumn{1}{c}{$X_7$}&
\multicolumn{1}{c}{$X_8$}&
\multicolumn{1}{c}{$X_9$}&
\multicolumn{1}{c}{$X_{10}$}&
\multicolumn{1}{c}{SB}\\
\hline
181558&A0& 12& 0.127& 0.808&   24.90& ---  & $<$1.00& 0.00  & 4.167& 4.16& --- &0\\
&& (8) & (8) & (8) & (8) & --- & (8) & (4) & (10) & (10) & --- & \\
112413&A0spe& 17& 0.183& ---  &$<$40.80& ---  &$<$40.80&$>$3.20& 4.061& 4.12& --- &0\\
&& (13) & (13) & --- & (2) & --- & (2) & (13) & (13) & (13) & --- & \\
\end{tabular}
\end{table}

\end{document}